\documentclass[twocolumn,letterpaper,aps,prc,longbibliography,superscriptaddress,nofootinbib,floatfix]{revtex4-1}

\usepackage{graphicx}	% Include figure files

\usepackage{amsmath}	% better math options
\usepackage{multirow}
\usepackage{amssymb}
\usepackage{bm}
\usepackage{float}
\usepackage{xspace}	% Include xspace

\begin{document}

\title{Measurement of charm and bottom production from semileptonic 
hadron decays in $p$$+$$p$ collisions at $\sqrt{s}=200$ GeV}

\newcommand{\abilene}{Abilene Christian University, Abilene, Texas 79699, USA}
\newcommand{\augie}{Department of Physics, Augustana University, Sioux Falls, South Dakota 57197, USA}
\newcommand{\banaras}{Department of Physics, Banaras Hindu University, Varanasi 221005, India}
\newcommand{\barc}{Bhabha Atomic Research Centre, Bombay 400 085, India}
\newcommand{\baruch}{Baruch College, City University of New York, New York, New York, 10010 USA}
\newcommand{\bnlcoll}{Collider-Accelerator Department, Brookhaven National Laboratory, Upton, New York 11973-5000, USA}
\newcommand{\bnlphys}{Physics Department, Brookhaven National Laboratory, Upton, New York 11973-5000, USA}
\newcommand{\caucr}{University of California-Riverside, Riverside, California 92521, USA}
\newcommand{\charlesczech}{Charles University, Ovocn\'{y} trh 5, Praha 1, 116 36, Prague, Czech Republic}
\newcommand{\chonbuk}{Chonbuk National University, Jeonju, 561-756, Korea}
\newcommand{\cns}{Center for Nuclear Study, Graduate School of Science, University of Tokyo, 7-3-1 Hongo, Bunkyo, Tokyo 113-0033, Japan}
\newcommand{\colorado}{University of Colorado, Boulder, Colorado 80309, USA}
\newcommand{\columbia}{Columbia University, New York, New York 10027 and Nevis Laboratories, Irvington, New York 10533, USA}
\newcommand{\czechtech}{Czech Technical University, Zikova 4, 166 36 Prague 6, Czech Republic}
\newcommand{\debrecen}{Debrecen University, H-4010 Debrecen, Egyetem t{\'e}r 1, Hungary}
\newcommand{\elte}{ELTE, E{\"o}tv{\"o}s Lor{\'a}nd University, H-1117 Budapest, P{\'a}zm{\'a}ny P.~s.~1/A, Hungary}
\newcommand{\eszterhazy}{Eszterh\'azy K\'aroly University, K\'aroly R\'obert Campus, H-3200 Gy\"ongy\"os, M\'atrai \'ut 36, Hungary}
\newcommand{\ewha}{Ewha Womans University, Seoul 120-750, Korea}
\newcommand{\famu}{Florida A\&M University, Tallahassee, FL 32307, USA}
\newcommand{\fsu}{Florida State University, Tallahassee, Florida 32306, USA}
\newcommand{\gsu}{Georgia State University, Atlanta, Georgia 30303, USA}
\newcommand{\hiroshima}{Hiroshima University, Kagamiyama, Higashi-Hiroshima 739-8526, Japan}
\newcommand{\howard}{Department of Physics and Astronomy, Howard University, Washington, DC 20059, USA}
\newcommand{\ihepprot}{IHEP Protvino, State Research Center of Russian Federation, Institute for High Energy Physics, Protvino, 142281, Russia}
\newcommand{\illuiuc}{University of Illinois at Urbana-Champaign, Urbana, Illinois 61801, USA}
\newcommand{\inrras}{Institute for Nuclear Research of the Russian Academy of Sciences, prospekt 60-letiya Oktyabrya 7a, Moscow 117312, Russia}
\newcommand{\instpasczech}{Institute of Physics, Academy of Sciences of the Czech Republic, Na Slovance 2, 182 21 Prague 8, Czech Republic}
\newcommand{\isu}{Iowa State University, Ames, Iowa 50011, USA}
\newcommand{\jaea}{Advanced Science Research Center, Japan Atomic Energy Agency, 2-4 Shirakata Shirane, Tokai-mura, Naka-gun, Ibaraki-ken 319-1195, Japan}
\newcommand{\kek}{KEK, High Energy Accelerator Research Organization, Tsukuba, Ibaraki 305-0801, Japan}
\newcommand{\korea}{Korea University, Seoul, 02841}
\newcommand{\kurchatov}{National Research Center ``Kurchatov Institute", Moscow, 123098 Russia}
\newcommand{\kyoto}{Kyoto University, Kyoto 606-8502, Japan}
\newcommand{\lawllnl}{Lawrence Livermore National Laboratory, Livermore, California 94550, USA}
\newcommand{\losalamos}{Los Alamos National Laboratory, Los Alamos, New Mexico 87545, USA}
\newcommand{\lund}{Department of Physics, Lund University, Box 118, SE-221 00 Lund, Sweden}
\newcommand{\lyon}{IPNL, CNRS/IN2P3, Univ Lyon, Université Lyon 1, F-69622, Villeurbanne, France}
\newcommand{\maryland}{University of Maryland, College Park, Maryland 20742, USA}
\newcommand{\mass}{Department of Physics, University of Massachusetts, Amherst, Massachusetts 01003-9337, USA}
\newcommand{\michigan}{Department of Physics, University of Michigan, Ann Arbor, Michigan 48109-1040, USA}
\newcommand{\muhlenberg}{Muhlenberg College, Allentown, Pennsylvania 18104-5586, USA}
\newcommand{\nara}{Nara Women's University, Kita-uoya Nishi-machi Nara 630-8506, Japan}
\newcommand{\natmephi}{National Research Nuclear University, MEPhI, Moscow Engineering Physics Institute, Moscow, 115409, Russia}
\newcommand{\newmex}{University of New Mexico, Albuquerque, New Mexico 87131, USA}
\newcommand{\nmsu}{New Mexico State University, Las Cruces, New Mexico 88003, USA}
\newcommand{\northcg}{Physics and Astronomy Department, University of North Carolina at Greensboro, Greensboro, North Carolina 27412, USA}
\newcommand{\ohio}{Department of Physics and Astronomy, Ohio University, Athens, Ohio 45701, USA}
\newcommand{\ornl}{Oak Ridge National Laboratory, Oak Ridge, Tennessee 37831, USA}
\newcommand{\orsay}{IPN-Orsay, Univ.~Paris-Sud, CNRS/IN2P3, Universit\'e Paris-Saclay, BP1, F-91406, Orsay, France}
\newcommand{\peking}{Peking University, Beijing 100871, People's Republic of China}
\newcommand{\pnpi}{PNPI, Petersburg Nuclear Physics Institute, Gatchina, Leningrad region, 188300, Russia}
\newcommand{\riken}{RIKEN Nishina Center for Accelerator-Based Science, Wako, Saitama 351-0198, Japan}
\newcommand{\rikjrbrc}{RIKEN BNL Research Center, Brookhaven National Laboratory, Upton, New York 11973-5000, USA}
\newcommand{\rikkyo}{Physics Department, Rikkyo University, 3-34-1 Nishi-Ikebukuro, Toshima, Tokyo 171-8501, Japan}
\newcommand{\saispbstu}{Saint Petersburg State Polytechnic University, St.~Petersburg, 195251 Russia}
\newcommand{\seoulnat}{Department of Physics and Astronomy, Seoul National University, Seoul 151-742, Korea}
\newcommand{\stonybrkc}{Chemistry Department, Stony Brook University, SUNY, Stony Brook, New York 11794-3400, USA}
\newcommand{\stonycrkp}{Department of Physics and Astronomy, Stony Brook University, SUNY, Stony Brook, New York 11794-3800, USA}
\newcommand{\tenn}{University of Tennessee, Knoxville, Tennessee 37996, USA}
\newcommand{\titech}{Department of Physics, Tokyo Institute of Technology, Oh-okayama, Meguro, Tokyo 152-8551, Japan}
\newcommand{\tsukuba}{Tomonaga Center for the History of the Universe, University of Tsukuba, Tsukuba, Ibaraki 305, Japan}
\newcommand{\vandy}{Vanderbilt University, Nashville, Tennessee 37235, USA}
\newcommand{\weizmann}{Weizmann Institute, Rehovot 76100, Israel}
\newcommand{\wigner}{Institute for Particle and Nuclear Physics, Wigner Research Centre for Physics, Hungarian Academy of Sciences (Wigner RCP, RMKI) H-1525 Budapest 114, POBox 49, Budapest, Hungary}
\newcommand{\yonsei}{Yonsei University, IPAP, Seoul 120-749, Korea}
\newcommand{\zagreb}{Department of Physics, Faculty of Science, University of Zagreb, Bijeni\v{c}ka c.~32 HR-10002 Zagreb, Croatia}
\affiliation{\abilene}
\affiliation{\augie}
\affiliation{\banaras}
\affiliation{\barc}
\affiliation{\baruch}
\affiliation{\bnlcoll}
\affiliation{\bnlphys}
\affiliation{\caucr}
\affiliation{\charlesczech}
\affiliation{\chonbuk}
\affiliation{\cns}
\affiliation{\colorado}
\affiliation{\columbia}
\affiliation{\czechtech}
\affiliation{\debrecen}
\affiliation{\elte}
\affiliation{\eszterhazy}
\affiliation{\ewha}
\affiliation{\famu}
\affiliation{\fsu}
\affiliation{\gsu}
\affiliation{\hiroshima}
\affiliation{\howard}
\affiliation{\ihepprot}
\affiliation{\illuiuc}
\affiliation{\inrras}
\affiliation{\instpasczech}
\affiliation{\isu}
\affiliation{\jaea}
\affiliation{\kek}
\affiliation{\korea}
\affiliation{\kurchatov}
\affiliation{\kyoto}
\affiliation{\lawllnl}
\affiliation{\losalamos}
\affiliation{\lund}
\affiliation{\lyon}
\affiliation{\maryland}
\affiliation{\mass}
\affiliation{\michigan}
\affiliation{\muhlenberg}
\affiliation{\nara}
\affiliation{\natmephi}
\affiliation{\newmex}
\affiliation{\nmsu}
\affiliation{\northcg}
\affiliation{\ohio}
\affiliation{\ornl}
\affiliation{\orsay}
\affiliation{\peking}
\affiliation{\pnpi}
\affiliation{\riken}
\affiliation{\rikjrbrc}
\affiliation{\rikkyo}
\affiliation{\saispbstu}
\affiliation{\seoulnat}
\affiliation{\stonybrkc}
\affiliation{\stonycrkp}
\affiliation{\tenn}
\affiliation{\titech}
\affiliation{\tsukuba}
\affiliation{\vandy}
\affiliation{\weizmann}
\affiliation{\wigner}
\affiliation{\yonsei}
\affiliation{\zagreb}
\author{C.~Aidala} \affiliation{\michigan} 
\author{Y.~Akiba} \email[PHENIX Spokesperson: ]{akiba@rcf.rhic.bnl.gov} \affiliation{\riken} \affiliation{\rikjrbrc} 
\author{M.~Alfred} \affiliation{\howard} 
\author{V.~Andrieux} \affiliation{\michigan} 
\author{N.~Apadula} \affiliation{\isu} 
\author{H.~Asano} \affiliation{\kyoto} \affiliation{\riken} 
\author{B.~Azmoun} \affiliation{\bnlphys} 
\author{V.~Babintsev} \affiliation{\ihepprot} 
\author{N.S.~Bandara} \affiliation{\mass} 
\author{K.N.~Barish} \affiliation{\caucr} 
% ...
\collaboration{PHENIX Collaboration} \noaffiliation

\date{\today}

%------------------------------------------------------------------------------|
\begin{abstract}

%\linenumbers

Measurements of the differential production of electrons from 
open-heavy-flavor hadrons with charm- and bottom-quark content in 
$p$$+$$p$ collisions at $\sqrt{s}=200$ GeV are presented.  The 
measurements proceed through displaced-vertex analyses of electron 
tracks from the semileptonic decay of charm and bottom hadrons using the 
PHENIX silicon-vertex detector.  The relative contribution of electrons 
from bottom decays to inclusive heavy-flavor-electron production is 
found to be consistent with fixed-order-plus-next-to-leading-log 
perturbative-QCD calculations within experimental and theoretical 
uncertainties.  These new measurements in $p$$+$$p$ collisions provide a 
precision baseline for comparable forthcoming measurements in A$+$A 
collisions.

\end{abstract}

\maketitle

%%%%%%%%%%%%%%%%%%%%%%%%%%%%%%%%%%%%%%%%%%%%%%%%%%%%%%%%%%%%%%%%%%%%%%%%%%%
\section{Introduction}
\label{sec:introduction}

Charm and bottom quarks are collectively referred to as heavy-flavor 
quarks.  Their production in elementary $p$$+$$p$ collisions is of 
interest from a variety of vantage points, both in high-energy particle 
and nuclear physics.

%The production of charm and bottom quarks---here collectively referred 
%to as heavy flavor---in elementary $p$$+$$p$ collisions is of interest 
%from a variety of vantage points, both in high-energy particle and 
%nuclear physics.

From a fundamental standpoint, unlike light quarks the large masses of 
heavy-flavor quarks (compare $m_c\approx 1280$ MeV/$c^2$ and 
$m_b\approx 4180$ MeV/$c^2$ with $m_u\approx 2.2$ MeV/$c^2$ and 
$m_d\approx 4.7$ MeV/$c^2$)~\cite{PhysRevD.98.030001} are such that 
their production can be calculated using perturbative quantum 
chromodynamics (pQCD) even at low $p_T$. At leading order (LO), heavy 
quark production proceeds via gluon fusion and quark-antiquark 
annihilation. At next-to-leading order (NLO), processes such as flavor 
excitation and gluon splitting are involved. In this regime, divergences 
are regulated by the mass of the heavy quarks, which acts as an infrared 
cutoff except when the quark $p_T$ is greater than its 
mass~\cite{Andronic:2015wma}. In that case, logarithmic divergences 
appear. The most advanced analytic pQCD techniques currently available 
allow for such divergences to be resummed, giving rise to the 
fixed-order-plus-next-to-leading-log (FONLL) 
approach~\cite{Cacciari:2005rk}. Unfortunately, FONLL calculations 
exhibit very large error bands associated predominantly with 
uncertainties in the heavy quark masses and the renormalization scales, 
motivating the need for comparisons with experimental data.

A wealth of heavy-flavor-production data exists both at the Relativistic 
Heavy Ion Collider (RHIC)~\cite{Averbeck:2013oga} and the Large Hadron 
Collider (LHC)~\cite{Andronic:2015wma}. At the LHC, such measurements 
comprise cross section measurements of inclusive heavy-flavor leptons, 
as well as of individual $D$ (containing charm) and $B$ (containing 
bottom) meson states.  At RHIC, such measurements are consistent with 
FONLL calculations within uncertainties, yet systematically higher than 
the central value predicted by the theory. It is thus of interest to 
arrive at a simultaneous measurement of charm and bottom production at 
RHIC energies to leverage the distinct masses of these quarks to provide 
constraints for pQCD calculations.

Now, from the standpoint of high energy nuclear physics, heavy-ion 
collisions at RHIC and the LHC produce deconfined nuclear matter---known 
as the quark-gluon plasma (QGP). The QGP produced in these colliders can be 
characterized as a strongly coupled fluid~\cite{RevModPhys.89.035001} 
exhibiting, among other properties, substantial color opacity. This 
refers to the ability of the medium to hinder the passage of color 
charges, resulting in the energy loss of such 
particles~\cite{Gyulassy:1990ye}. Charm and bottom quarks are excellent 
probes of color opacity because they originate primarily from 
early-stage hard-parton-scattering processes, and thus transit through 
the entire evolution of the QGP medium~\cite{Averbeck:2013oga}.

The yield of heavy-flavor electrons at RHIC scales with the number of 
binary nucleon-nucleon collisions~\cite{Adler:2004ta,Adare:2006nq} as a 
consequence of charm and bottom conservation by the strong interaction. 
Nevertheless, their spectrum is modified in central Au$+$Au collisions 
relative to the $p$$+$$p$ baseline, as quantified by the nuclear 
modification factor $R_{AA}$~\cite{Adare:2006nq}. Heavy quarks are 
redistributed in momentum space, such that a strong suppression of 
heavy-flavor electrons is observed for $p_T>5$ GeV/$c$, comparable in 
magnitude to that observed for light 
quarks~\cite{Adcox:2001jp,Adler:2003qi}.

This constitutes a puzzling observation, as it challenges traditional 
interpretations of energy loss as proceeding exclusively through gluon 
radiation, requiring the inclusion of additional collisional mechanisms. 
To shed light on the interplay of radiative and collisional energy loss 
by leveraging the mass difference between charm and bottom, the PHENIX 
collaboration has measured separated heavy-flavor-quark yields from 
semileptonic decay electrons in Au$+$Au collisions using the 
silicon-vertex-detector upgrade~\cite{Adare:2015hla}. Nuclear 
modification factors $R_{AA}$ were calculated using a $p$$+$$p$ baseline 
measurement by the STAR collaboration~\cite{Aggarwal:2010xp} obtained 
via electron-hadron correlations with limited kinematic reach and large 
uncertainties.

In this paper, we present a new baseline measurement of heavy-flavor 
separation in $p$$+$$p$ at $\sqrt{s}=200$ GeV using the same 
displaced-vertex analysis technique used in a previous PHENIX 
measurement made in Au$+$Au collisions at $\sqrt{s}=200$ 
GeV~\cite{Adare:2015hla}.  Our new results with smaller uncertainties 
and extended kinematic range provide a valuable update for future 
measurements of heavy-flavor modification

\section{Experimental Setup}
\label{sec:exp_setup}

%%%%%%%%%%%%%  FIG: PHENIX schematic configuration %%%%%% Fig_1
\begin{figure*}[htb]
\includegraphics[width=0.7\linewidth]{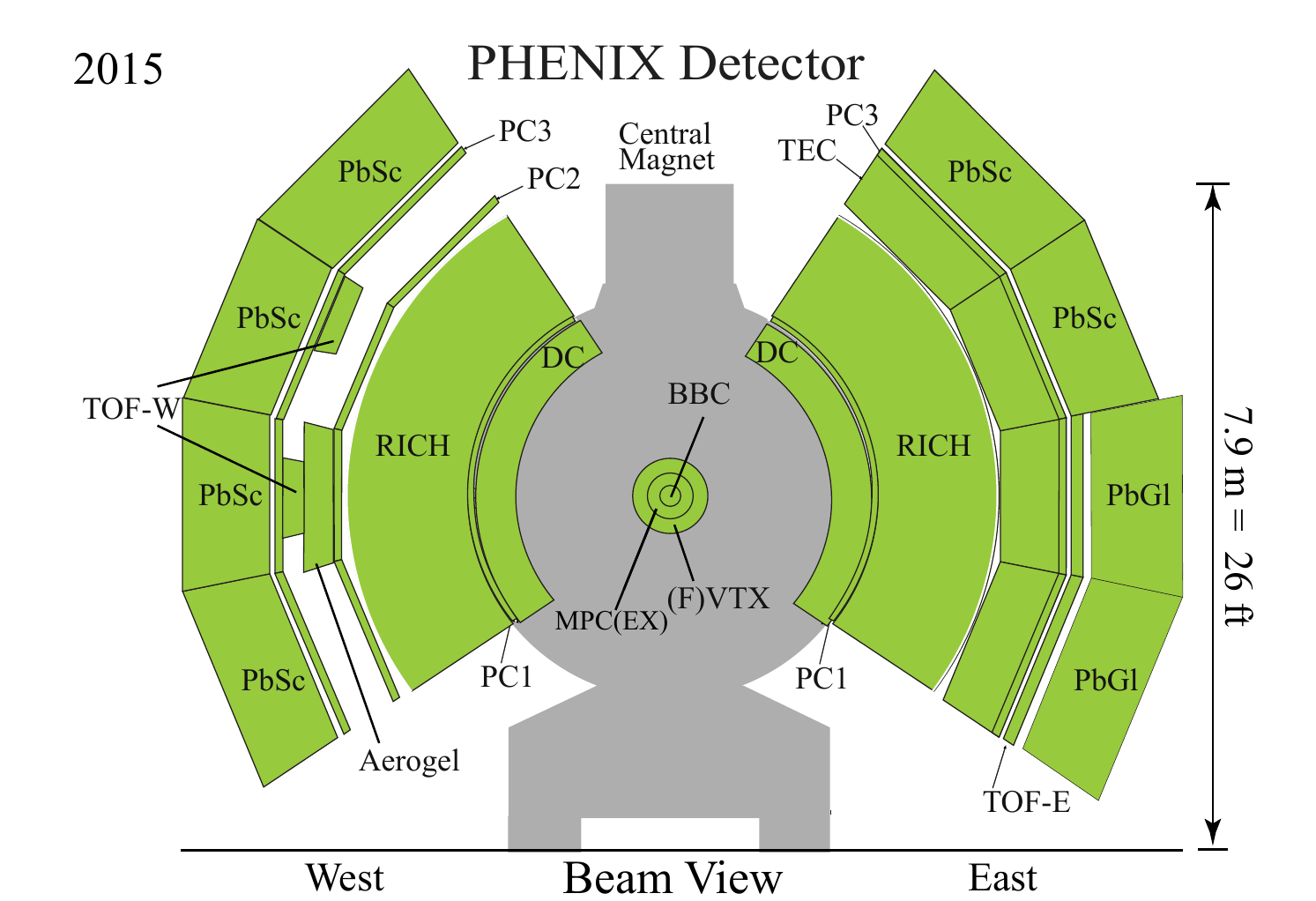}
\caption{Schematic view of the PHENIX detector configuration along the 
beam axis in the 2015 run period.  The indicated subsystems are 
identified in the text.}
\label{fig_phenix}
\end{figure*}

Figure~\ref{fig_phenix} shows a transverse (beam) view of the PHENIX 
detector and its subsystems. Two midrapidity spectrometers, called the 
central arms, are shown on either side of the central magnet. With an 
acceptance of $|\eta| < 0.35$ and $\Delta\phi = \pi/2$, each arm 
provides tracking and particle identification capabilities. The magnetic 
field is generated by two pairs of coils in the pole faces of the 
central magnet such that when electric current runs in the same 
direction in both coils, a maximum field strength of 0.9 T is achieved 
at the beam location. A detailed description of the PHENIX detector is 
given in Ref.~\cite{ADCOX2003469,mannel_vtx,osti_15007383}.

The drift chambers and three layers of multiwire-proportional pad 
chambers are the subsystems used for charged particle 
tracking~\cite{Adcox:2003zp}. The Ring Imaging \v{C}erenkov (RICH) 
detector and the electromagnetic calorimeter (EMCal) are the subsystems 
used for electron identification. The RICH~\cite{Aizawa:2003zq} 
comprises two independent volumes, one in each detector arm, filled with 
CO$_2$. The gas acts as a dielectric medium, in which electrons emit 
\v{C}erenkov radiation for $p_T>20$ MeV/$c$; pions can also emit light 
in the RICH above $p_T\approx5$ GeV/$c$. The 
EMCal~\cite{Aphecetche:2003zr}, which comprises lead-glass (PbGl) and 
lead-scintillator (PbSc) modules, is used to identify electrons based on 
the transverse shape of an electromagnetic shower and the ratio of the 
particle's energy deposit in the EMCal to the momentum of the 
reconstructed track.

%%%%%%%%%%%%%%  FIG: VTX detector %%%%%%%%%%%%%%%%%%%%%%% Fig_2
\begin{figure}[htb]
\includegraphics[width=1.0\linewidth]{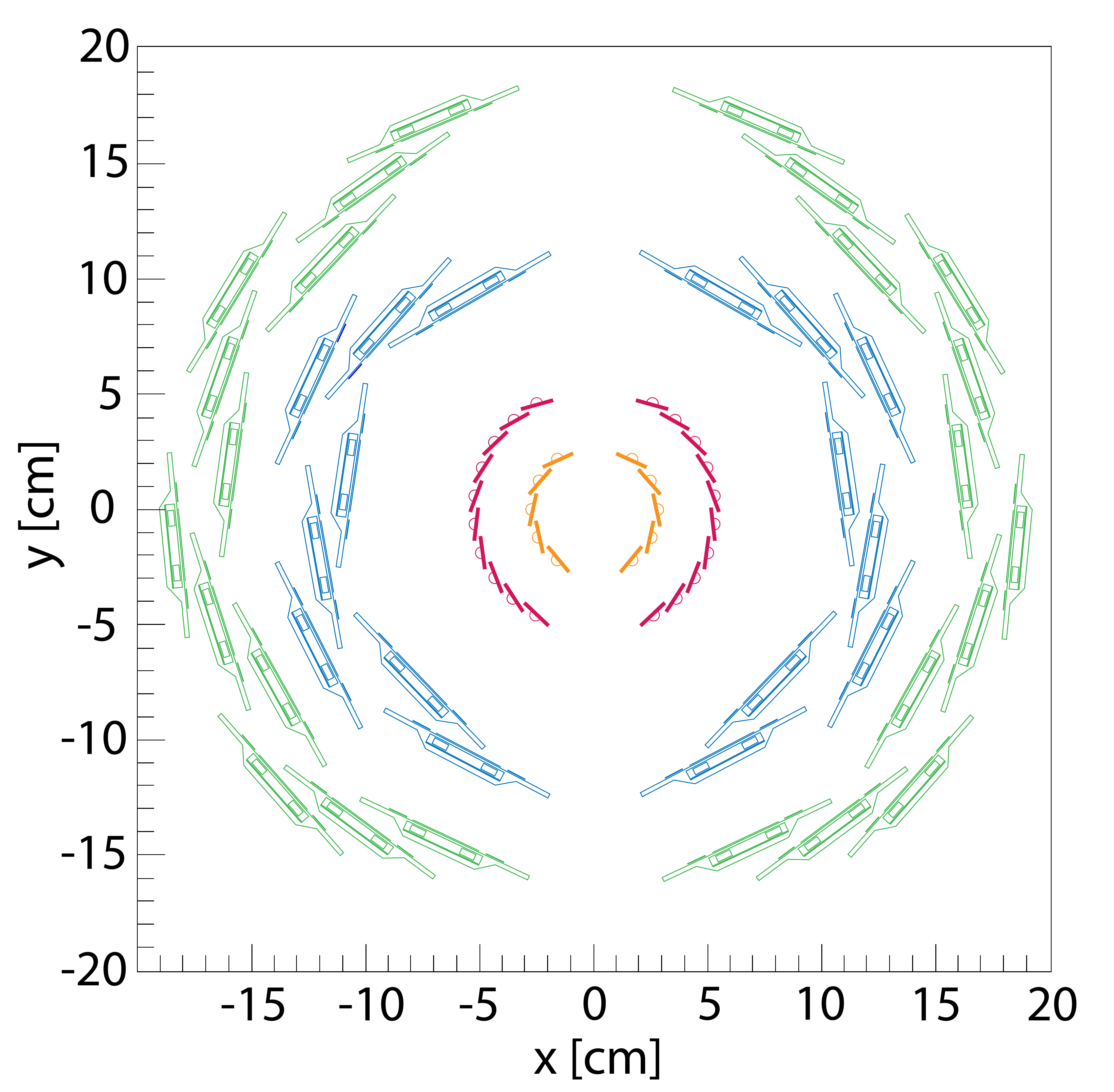}
\caption{Cross sectional view of the VTX detector showing the relative 
positions of individual layers B0, B1, B2, and B3 from smallest 
to largest radius.}
\label{fig_vtx}
\end{figure}

Figure~\ref{fig_vtx} shows the finely segmented silicon-vertex detector 
(VTX)~\cite{mannel_vtx,1748-0221-4-04-P04011}, which was installed as an 
upgrade in 2011 to provide tracking close to the interaction region, 
capable of reconstructing the primary vertex with a resolution on the 
order of 150 $\mu$m in $p$$+$$p$ collisions.  The VTX comprises two arms 
with four independent layers arranged around the beam pipe at nominal 
radii of $r=2.6$, 5.1, 11.8, and 16.7 cm. The material budget, expressed 
as a percentage of a radiation length is, for each layer, 
$X_0(\%)=1.28$, 1.28, 5.43, and 5.43. Simultaneously with the VTX a new, 
thinner, beryllium beam pipe was installed in 2011 with a material 
budget of $X_0(\%)=0.22$.  Each layer comprises a series of ladders 
extending longitudinally. The VTX has an acceptance of $|\eta| < 1$ and 
$\Delta\phi \approx 0.8\pi$ per arm. Going from smallest to largest 
radius, individual layers are named B0, B1, B2, and B3. The innermost 
layers, B0 and B1, were constructed using silicon-pixel technology 
developed at CERN~\cite{Ichimiya:2008am}. Pixels in these layers have 
dimensions $50\mu {\rm m}\times 425 \mu {\rm m}$, and are arranged into 
lattices of $256\times 32$ pixels which are read out by a single 
ALICE1LHCb sensor-readout chip~\cite{Snoeys:2000hn}. Four readout chips 
constitute one sensor module, with four sensor modules in a single 
ladder. Layers B0 and B1 have 5 and 10 ladders per arm, respectively. 
Layers B2 and B3, were constructed using a novel silicon-stripixel 
technology developed at Brookhaven National Laboratory.  Each 
$4.34\times6.46$ cm sensor in these layers is segmented into 
$80\mu{\rm m}\times 1000\mu{\rm m}$ stripixels. These are implanted with two 
serpentine metal strips defining two readout directions, $X$ and $U$, 
such that the two-dimensional location of hit positions can be 
determined. Layers B2 and B3 have 8 and 12 ladders per arm, 
respectively, with 5(6) sensors per ladder in B2(B3).  Stripixel sensors 
are read out using the SVX4 readout chip, developed by a collaboration 
between Fermilab and Lawrence Berkeley National 
Laboratory~\cite{Krieger:2003jr}.

%%%%%%%%%%%%%%%%%%%%%%%%%%%%%%%%%%%%%%%%%%%%%%%%%%%%%%%%%%%%%%%%%%%%%%%%%%%
\section{Methods}

The goal of this analysis is to measure the invariant yield of 
heavy-flavor electrons, independently for charm and bottom decays. This 
is accomplished by exploiting the fact that hadrons with bottom content 
have a longer lifetime than those with charm, as shown in 
Table~\ref{table_lifetimes} for $B$ and $D$ 
mesons~\cite{PhysRevD.98.030001}. As will be described in the following 
subsection, the provenance of heavy flavor electron tracks is determined 
statistically based on the distance of closest approach in the 
transverse plane (DCA$_T$) between the tracks and the beam center, which 
is the point relative to which they are reconstructed,

%======================================================  Table_I
\begin{table}
\caption{Lifetime $c\tau_0$ of selected $D$ and $B$ 
states~\protect\cite{PhysRevD.98.030001}.}
\begin{ruledtabular} \begin{tabular}{cccccc}
\hline
&& Particle & Lifetime $c\tau_0$ && \\
\hline
&& $D^0$ & 129.9 $\mu$m && \\
&& $D^+$ & 311.8 $\mu$m && \\
\\
&& $B^0$ & 457.2 $\mu$m&& \\
&& $B^+$ & 491.1 $\mu$m && \\
\end{tabular} \end{ruledtabular}
\label{table_lifetimes}
\end{table}

Thus, the longer lifetime of the $B$, and its decay kinematics, will 
result in a broader DCA$_T$ distribution than for electrons from the 
shorter-lived $D$ mesons. However, the measured electron candidate 
sample contains not only heavy flavor electrons, but also abundant 
background from a variety of sources (i.e., decays of $\pi^0$, $\eta$, 
$\rho$, $\omega$ $J/\psi$, $K^{\pm}$, $K_s^0$, $\Upsilon$ mesons and the 
Drell-Yan process, as well as conversions of direct and decay photons), 
each with its own characteristic DCA$_T$ shape. Once this background has 
been determined, the DCA$_T$ distribution of inclusive heavy flavor 
electrons can be isolated.  The individual contributions from charm and 
bottom can then be obtained through an inversion procedure often 
referred to as unfolding~\cite{Choudalakis:2012hz}.  We outline the steps 
involved in the analysis as follows:

\begin{enumerate}

\item Measure the DCA$_T$ distribution of hadrons and electrons 
candidate tracks in data, as a function of track $p_T$.

\item Model the DCA$_T$ distributions of nonheavy-flavor background in 
the candidate electron sample by simulating the following electron 
sources: $\pi^0$, $\eta$, direct photons, $J/\psi$, $K_s^0$, $K^{\pm}$, 
and hadron contamination.

\item Determine the fraction of electrons attributable to each of the 
background sources considered, thus normalizing the background DCA$_T$ 
distributions relative to those of electron candidates in data.

\item Separate the contribution of charm and bottom decays to the 
electron sample using Bayesian inference techniques. This step is 
constrained by the measured electron DCA$_T$ distributions, as well as 
by the invariant yield of \textit{inclusive} heavy-flavor electrons, 
previously published by the PHENIX collaboration~\cite{Adare:2010de}.

\end{enumerate}

This analysis used 110 pb$^{-1}$ of integrated luminosity collected 
during the 2015 $p$$+$$p$ RHIC running period. A family of EMCal-RICH 
triggers were used to maximize the number of electron tracks available 
for analysis. These triggers segment the calorimeter and RICH detector 
into a series of tiles, triggering on events in which a certain energy 
threshold is exceeded in a calorimeter tile, and for some triggers 
requiring that a spatial match can be found in the RICH.

\subsection{Measuring Track DCA$_T$}

Track reconstruction is carried out using the central arm spectrometers, 
as detailed in Ref.~\cite{Adare:2006nq}. Electron candidates within $1.5 
< p_{T} {\rm [GeV/\textit{c}]} < 6.0$ are identified by matching 
reconstructed tracks with hits in the RICH, and energy deposits in the 
EMCal.

Electrons traversing the RICH emit \v{C}erenkov light, which is 
amplified by photomultiplier tubes (PMT). A maximum displacement of 5 cm 
is allowed between a track projection and the centroid of the hit PMTs. 
For tracks with $p_T < 5$ GeV/$c$, at least one PMT hit is required in 
the RICH, whereas at higher $p_T$ at least three hits are required, 
given that pions in this kinematic region begin to radiate in the RICH.

Additionally, the energy $E$ deposited by a track in the EMCal is 
required to match its momentum $p$, since---unlike hadrons---electrons 
deposit the majority of their energy in the calorimeter. This is 
quantified through the variable $dep=(E/p-\mu_{E/p})/\sigma_{E/p}$, 
where $\mu_{E/p}$ and $\sigma_{E/p}$ correspond to the mean and width of 
a Gaussian fit to the distribution of the energy-momentum ratio $E/p$ 
around $E/p = 1$, respectively. A cut on $|dep|<2$ is then used to 
select electrons.

Additional cuts involving the EMCal include restricting the displacement 
in $\Delta z$ and $\Delta \phi$ between the track projection and the 
calorimeter shower to within three standard deviations. Finally, a cut 
on the probability that a given EMCal cluster originates from an 
electromagnetic shower---as determined from the shower shape---is used 
to reject hadrons.

Once identified in the central arms, reconstructed tracks are projected 
back to the VTX detector, where an iterative algorithm described in 
Ref.~\cite{Adare:2015hla} is used to associate the track with VTX hits 
to create a VTX-associated track.  Such tracks are required to have a hit 
in each of the two innermost layers of the VTX, and at least one hit in 
either of the outer layers, and to satisfy $\chi^2_{{\rm vtx}}/{\rm ndf} 
< 2$ to ensure the quality of the fit.

%%%%%%%%%%%%%%%%%%%%%%%%%%%%%%%%%%%%%%%%%%%%%%%%%%  Fig_3
\begin{figure}[htb]
\includegraphics[width=1.0\linewidth]{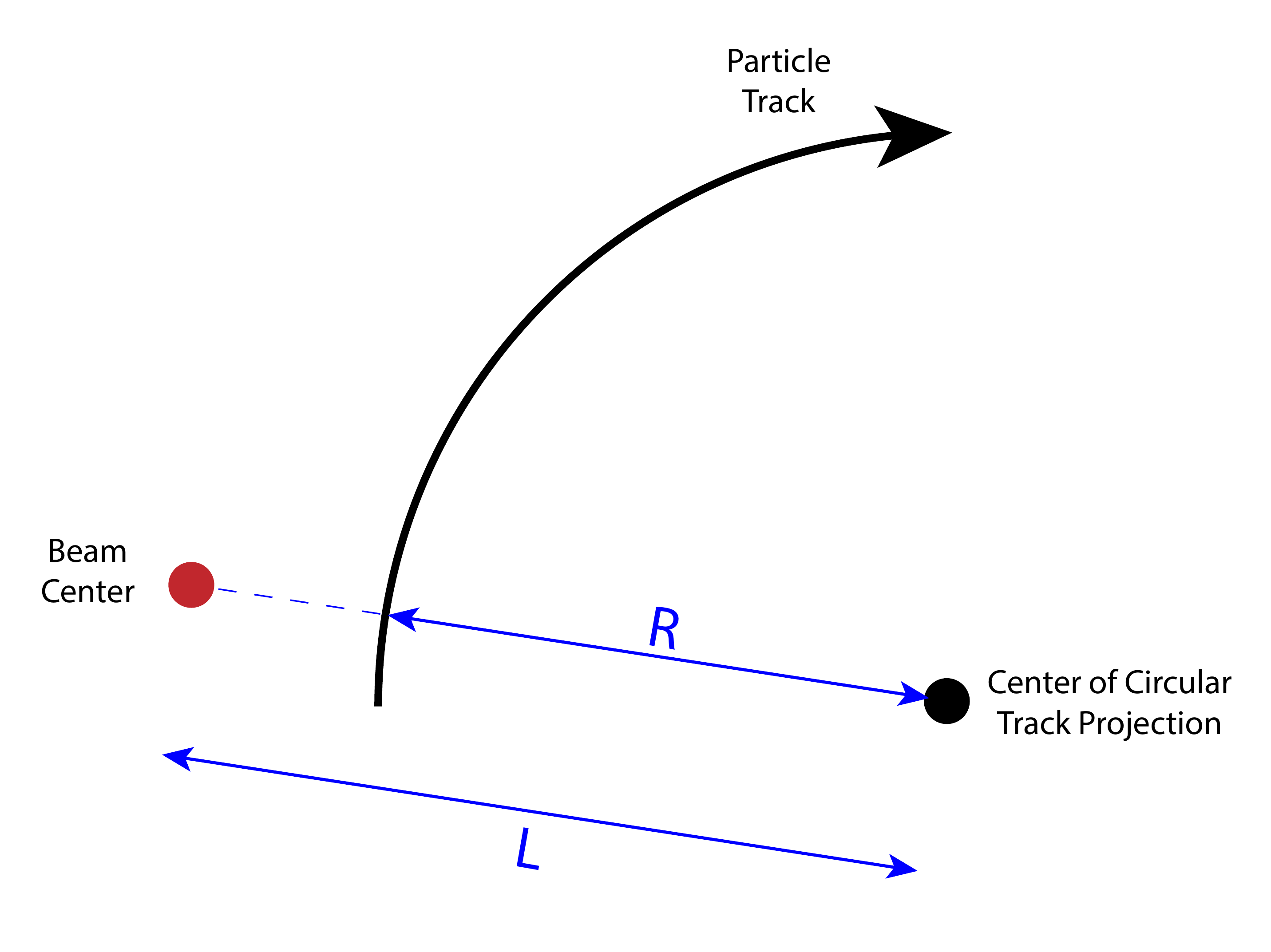}
\caption{Schematic diagram illustrating the definition of track DCA$_T$ 
in the transverse plane, as DCA$_T = L-R$.}
\label{fig_dca_schematic}
\end{figure}

Figure~\ref{fig_dca_schematic} shows a schematic diagram defining the 
DCA$_T$ of a VTX-associated track.  The circular track projection is 
shown in the transverse plane, where a constant magnetic field exists 
over the region covered by the VTX detector. The DCA$_T$ is then defined 
as ${\rm DCA}_T=L-R$, where $L$ is the distance between the beam center 
and the center of the projection, and $R$ is the projection radius. The 
beam center is defined as the geometric center of the transverse region 
over which beam collisions occur, is constant over a given run period, 
and exhibits a Gaussian spread of width 
$\sigma_x^{{\rm (beam)}}\approx130$ $\mu{\rm m}$ and $\sigma_y^{{\rm 
(beam)}}\approx100$ $\mu{\rm m}$.  Notice that DCA$_T$ is a signed 
quantity which is not generally symmetric around zero, because electrons 
from some background sources exhibit asymmetric DCA$_T$ distributions 
depending on the decay kinematics of their parent particles.

In a previous PHENIX analysis~\cite{Adare:2015hla}, the DCA$_T$ was 
defined relative to the primary vertex of the collision, rather 
than the beam center. The primary vertex is determined using tracks 
reconstructed from VTX hits alone, with no reliance on the central arm 
tracking subsystems. However, given the low multiplicity of $p$$+$$p$ 
collisions, such a procedure does not converge to a vertex for 
approximately 50\% all $p$$+$$p$ events. Furthermore, when it does 
converge---and particularly in events with electron tracks---the low 
number of reconstructed tracks makes it likely that the primary vertex 
is biased towards a displaced vertex and thus unsuitable for analysis. 
The choice of using the beam center for DCA$_T$ determination is further 
justified because the primary-vertex resolution is quite similar to the 
beam spot spread. Table~\ref{table_vertex_res} shows the resolution of 
the precise vertex in the transverse plane, as a function of the number 
of reconstructed VTX tracks used in its determination. The resolution 
improves significantly with increasing number of tracks. However, due to 
the limited coverage of the VTX, the average number of tracks used to 
calculate the primary vertex in $p$$+$$p$ collisions is 3.2, such that 
the corresponding resolution is broader than the beam spread.

%======================================================  Table_II
\begin{table}
\caption{Resolution of the primary vertex in the transverse plane, as a 
function of the number of reconstructed VTX tracks available for its 
determination.}
\begin{ruledtabular} \begin{tabular}{ccccc}
& Number    & $\sigma_x^{{\rm vertex}}$ & $\sigma_y^{{\rm vertex}}$ & \\
& of Tracks &        $[\mu{\rm m}]$     &       $[\mu{\rm m}]$      & \\
\hline
& 2 & 296.5 & 207.0 &\\
& 3 & 195.0 & 141.2 &\\
& 4 & 157.3 & 113.7 &\\
& 5 & 132.7 & 97.0  &\\
& 6 & 118.8 & 77.8  &\\
& 7 & 98.8  & 72.8  &\\
& 8 & 89.6  & 60.3  &\\
\end{tabular} \end{ruledtabular}
\label{table_vertex_res}
\end{table}

%%%%%%%%%%%%%%%%%%%%%%%%%%%%%%%%%%%%%%%%%%%%%%%%%%  Fig_4
\begin{figure}[htb]
\includegraphics[width=1.0\linewidth]{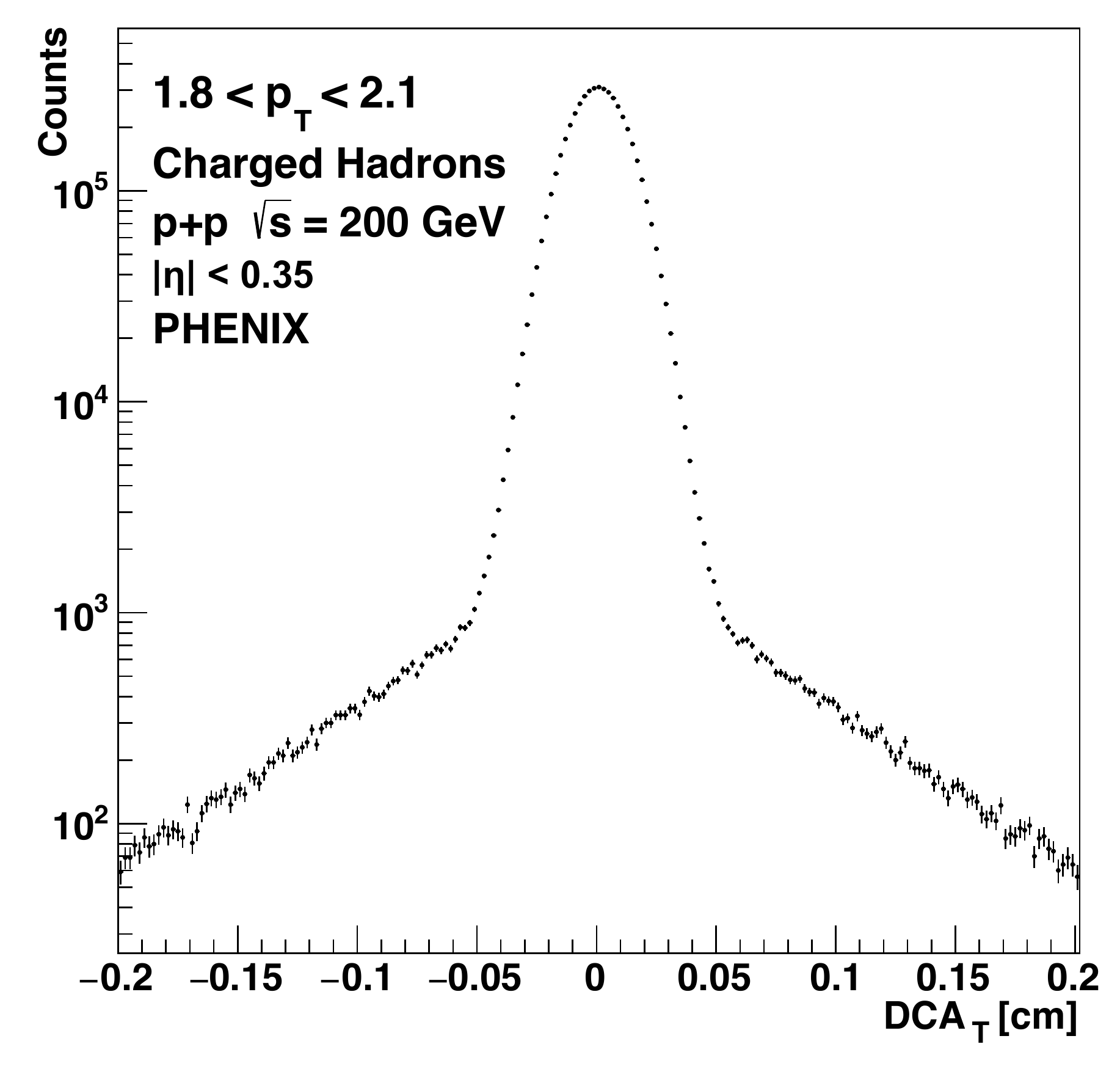}
\caption{DCA$_T$ distribution of hadron tracks in data within $1.8 {\rm 
GeV}/c< p_T<2.1 {\rm GeV}/c$, calculated relative to the beam center 
position.}
\label{fig_dca_hadron}
\end{figure}

Figure~\ref{fig_dca_hadron} shows the DCA$_T$ distribution of hadron 
tracks within $1.8 < p_T {\rm [GeV}/c] < 2.1$, where the histogram 
corresponds to counts of tracks passing the analysis cuts, with no 
correction for acceptance or efficiency effects. Hadron tracks are 
subject to the same quality requirements as electron tracks, but are 
identified as hadrons by requiring no RICH PMT hits. The very prominent 
Gaussian peak centered at ${\rm DCA}_T=0$ is attributed to particles 
originating from the primary collision point, with its width reflecting 
the beam spread, convolved with the track-pointing resolution. On the 
other hand, the broad tails can be attributed to long-lived light 
hadrons which decay, as well as background. It is observed that the 
DCA$_T$ resolution improves with increasing track $p_T$. For this 
analysis, DCA$_T$ distributions of electron candidate tracks were 
measured in 10 $p_T$ bins between $1.5 < p_T {\rm [GeV}/c] < 6.0$.

\subsection{Modeling Electron Background Sources}

In addition to heavy-flavor-decay electrons, the electron sample 
determined by applying the track cuts described in the previous section 
contains contributions from a variety of background sources. Namely, we 
consider $(i)$ photonic electrons from the Dalitz decay of $\pi^0$ and 
$\eta$ mesons, as well as photon conversions; $(ii)$ nonphotonic 
electrons from the decay of $J/\psi$ and the three-body decay of 
$K^{\pm}$ and $K_s^0$ (collectively called $Ke3$ electrons); and $(iii)$ 
hadrons which are misidentified as electrons.

To isolate the heavy-flavor signal of interest, it is necessary to 
properly account for the background. Conversion electrons constitute the 
single largest source of background in this analysis owing to the 
material budget of the the VTX with $X_0(\%)=13.42$ of a radiation 
length. In the following subsection we describe a strategy that uses the 
fine segmentation of the VTX itself to reject the vast majority of 
conversions based on the narrow opening angle topology of conversion 
electron pairs. The remaining background, both photonic and nonphotonic, 
is accounted for by constructing an electron cocktail normalized 
relative to the measured electron sample.

Of the background electron sources, all but misidentified hadrons can be 
modeled using previous measurements of primary (i.e., $\pi^0$, $\eta$, 
$J/\psi$, $K^{\pm}$,$K^0_s$) particle production combined with a 
knowledge of their decay modes and \textsc{geant3} simulations of the 
PHENIX detector. Contributions from other sources of electrons, like the 
decay of vector mesons such as the $\Upsilon$, $\phi$, $\omega$ and 
$\rho$, as well as the Drell-Yan process, were found to contribute 
negligibly to the total electron background in the kinematic region of 
interest.

\subsubsection{Photonic Electron Background}

Photonic electrons originate from the Dalitz decay ($X\rightarrow 
e^+e^-\gamma$) of $\pi^0$ and $\eta$ mesons, and from the conversion of 
photons ($\gamma\rightarrow e^+e^-$) interacting with the beam pipe or 
the VTX detector itself, where the photons are either direct photons or 
a hadronic decay product. To model this background, we start with the 
published cross section of $\pi^0$, $\eta$ and direct photons in 
$p$$+$$p$ at $\sqrt{s}=200$ 
GeV~\cite{Adare:2008ab,Adare:2008qb,Adler:2006yt,Adare:2007dg}. Single 
particles are then generated between $0 < p_{T}{\rm [GeV/}c] < 20$ 
according to the published spectrum. For $\pi^0$ and $\eta$, accounting 
for branching ratios, the decay is forced to proceed exclusively through 
channels involving photons or electrons in the final state. The decay 
photons and electrons are fed through a \textsc{geant3} simulation of 
the PHENIX detector, where the same reconstruction code and track cuts 
used in data are applied. The resulting reconstructed electron yield is 
normalized by the number of simulated primary particles, thus correctly 
describing the relative contribution of each primary source to the total 
photonic electron yield.

As previously mentioned, conversion electrons constitute the most 
significant source of background in this analysis, originating from the 
beam pipe as well as all four layers of the VTX. We can eliminate 80\% 
of conversion electrons by imposing the requirement that tracks used in 
analysis have a hit in each of the innermost two layers of the VTX, 
thereby discarding electron tracks originating in the outer layers.

Given the narrow opening angle between the $e^+e^-$ pair from photon 
conversions, a veto cut is defined to minimize the remaining conversions 
from the beam pipe and innermost two layers. In this approach, tracks 
with a VTX hit in close proximity, within a certain window in 
$\Delta\phi$ and $\Delta z$, are rejected. As illustration, if a 
conversion occurs in the beam pipe, or in B0, then nearby pairs of hits 
will be found in subsequent detector layers. If at least one of these 
electrons is reconstructed as a track, the conversion veto cut will 
reject it based on the presence of at least one nearby hit within the 
window in any layer.

%%%%%%%%%%%%%%%%%%%%%%%%%%%%%%%%%%%%%%%%%%%%%%%%%%  Fig_5
\begin{figure*}[tbh]
\includegraphics[width=0.99\linewidth]{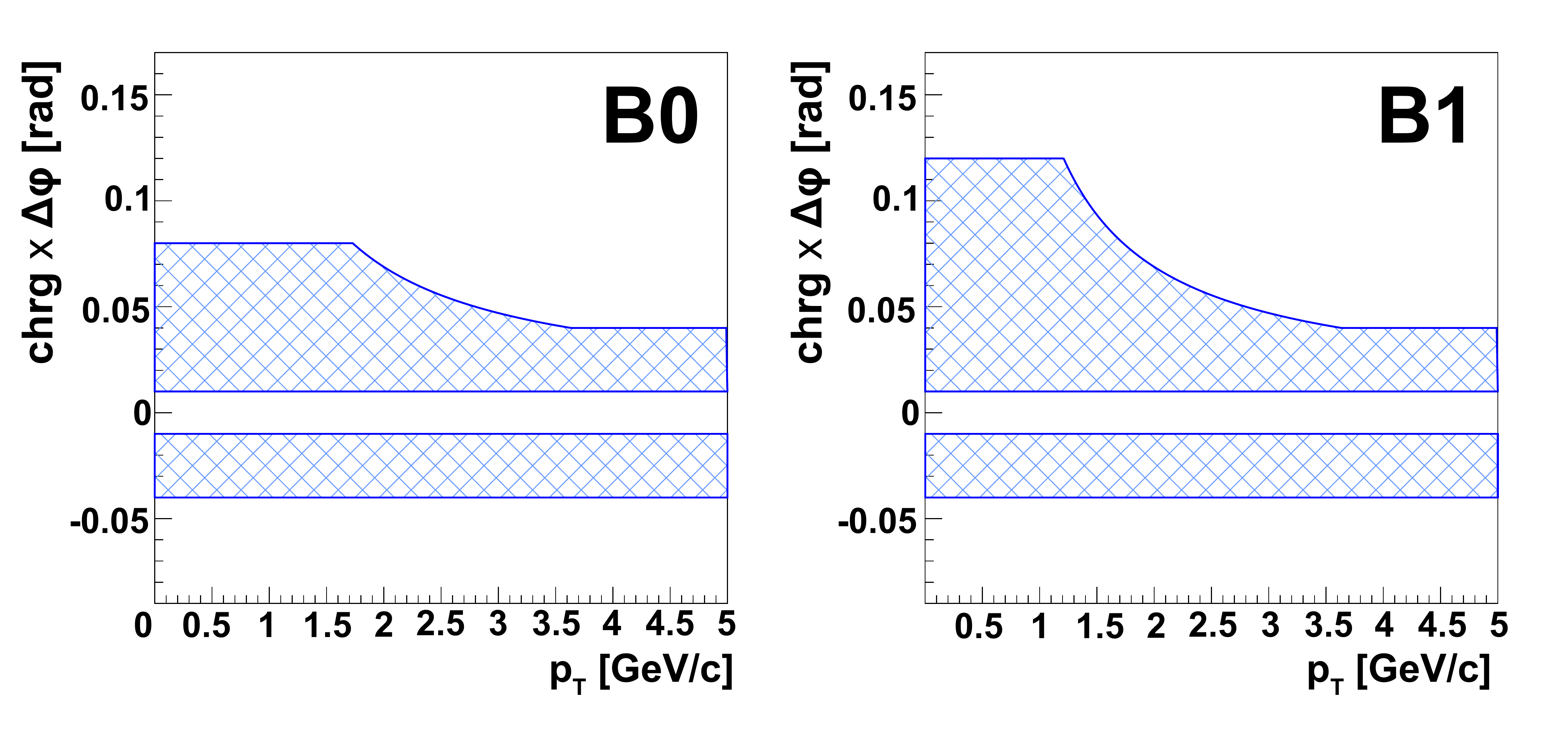}
\caption{Track $p_T$-dependent window in ${\rm chrg}\times\Delta\phi$ used in 
the conversion veto cut in the pixel detector layers, where chrg is 
the charge of a given track and $\Delta\phi$ is the azimuthal separation 
between clusters, one of which is associated with the track.}
\label{fig_veto_windows}
\end{figure*}

The size of the conversion veto window in ${\rm chrg}\times\Delta\phi$, 
where chrg is the charge of the track and $\Delta\phi$ is the azimuthal 
distance between track and cluster on the surface of a VTX chip, is 
shown in Fig.~\ref{fig_veto_windows}. It depends on the track $p_T$, as 
well as the layer where the nearby cluster is found. In general, because 
the bend of conversion electron pairs in the magnetic field decreases 
with photon momentum, the windows become narrower with increasing 
electron $p_T$. Furthermore, due to multiple scattering as well as the 
separation of electron pairs in the magnetic field, windows in the outer 
layers are larger than in the inner layers. The windows are asymmetric 
because the quantity ${\rm chrg}\times\Delta\phi$ is positive by 
construction; the negative side of the window is populated by 
mismeasured tracks which do not yield a positive ${\rm 
chrg}\times\Delta\phi$. In the longitudinal direction, the conversion 
veto window is $|\Delta z| < 0.05$ cm in the innermost two VTX layers, 
and $|\Delta z| < 0.1$ cm in the outermost two.

%%%%%%%%%%%%%%%%%%%%%%%%%%%%%%%%%%%%%%%%%%%%%%%%%%  Fig_6
\begin{figure}[htb]
\includegraphics[width=1.0\linewidth]{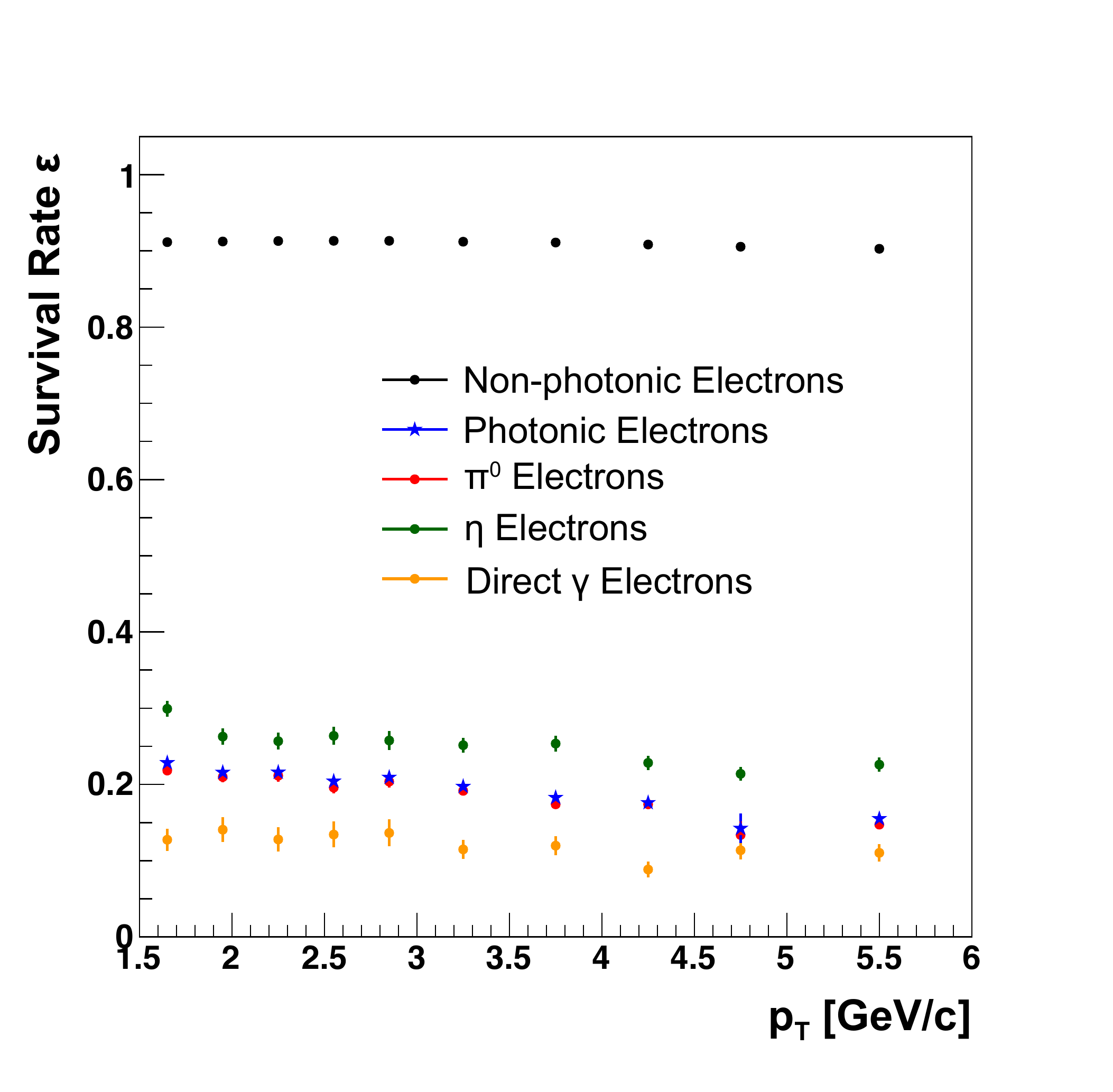}
\caption{Conversion veto cut survival rates for nonphotonic, and 
photonic electrons from various sources. The total photonic survival 
rate is the weighted average of the individual sources.}
\label{fig_surv_rates}
\end{figure}

The survival rate $\varepsilon$ of electrons from a given source is 
defined as the probability that they will not be rejected by the 
conversion veto cut.  Figure~\ref{fig_surv_rates} shows the survival 
rate of electrons from photonic and nonphotonic sources as a function of 
electron track $p_T$, where the nonphotonic survival rate has been 
estimated using hadrons in data as a proxy. The survival rate of 
photonic electrons has been further broken down by background source, 
namely $\pi^0$ and $\eta$ decays, as well as photon conversions.

Conversion electrons, such as those shown from direct photons, have the 
lowest survival rate of all.  Electrons from $\pi^0$ 
and $\eta$ mesons have a higher survival probability because they 
include---in addition to photon conversions---Dalitz electron pairs 
which have a wider opening angle than conversions pairs.  The survival 
rate of all photonic electrons combined is shown in blue in 
Fig.~\ref{fig_surv_rates} to be approximately 20\%. This demonstrates 
the ability of the conversion veto cut to reject a substantial fraction 
of the photonic background. In contrast, the survival rate of 
nonphotonic electrons is very high, at approximately 90\%. The 
conversion veto cut rejects a small fraction of nonphotonic electrons 
due to the presence of uncorrelated random hits in the window, which 
affect all tracks regardless of their provenance. The particular size of 
the conversion veto windows used in this analysis represents a 
compromise between maintaining a large window for background rejection, 
and limiting its size to minimize the inclusion of uncorrelated hits.

After applying the conversion veto cut on electrons in the photonic 
cocktail, the contribution of each primary particle source to the total 
photonic background can be calculated, as shown in 
Fig.~\ref{fig_cocktail}(a). Electrons from $\pi^0$ (both Dalitz and from 
the conversion of decay photons) dominate the background for all $p_T$, 
with direct photon external conversions becoming more significant at 
higher $p_T$.

\subsubsection{Non-photonic Electron Background}

Non-photonic electrons in this analysis correspond to those from the 
decay of $J/\psi$ mesons, as well as the three-body decays of $K^{\pm}$ 
and $K_s^0$, collectively known as $Ke3$ electrons 
($K{\rightarrow} e \nu \pi$). Other background electron sources, namely 
the decays of vector mesons such as the $\Upsilon$, $\rho$, $\omega$, 
and $\phi$ were considered in the background cocktail, but were found to 
contribute negligibly.

%%%%%%%%%%%%%%%%%%%%%%%%%%%%%%%%%%%%%%%%%%%%%%%%%%  Fig_7
\begin{figure*}[tbh]
\includegraphics[width=0.99\linewidth]{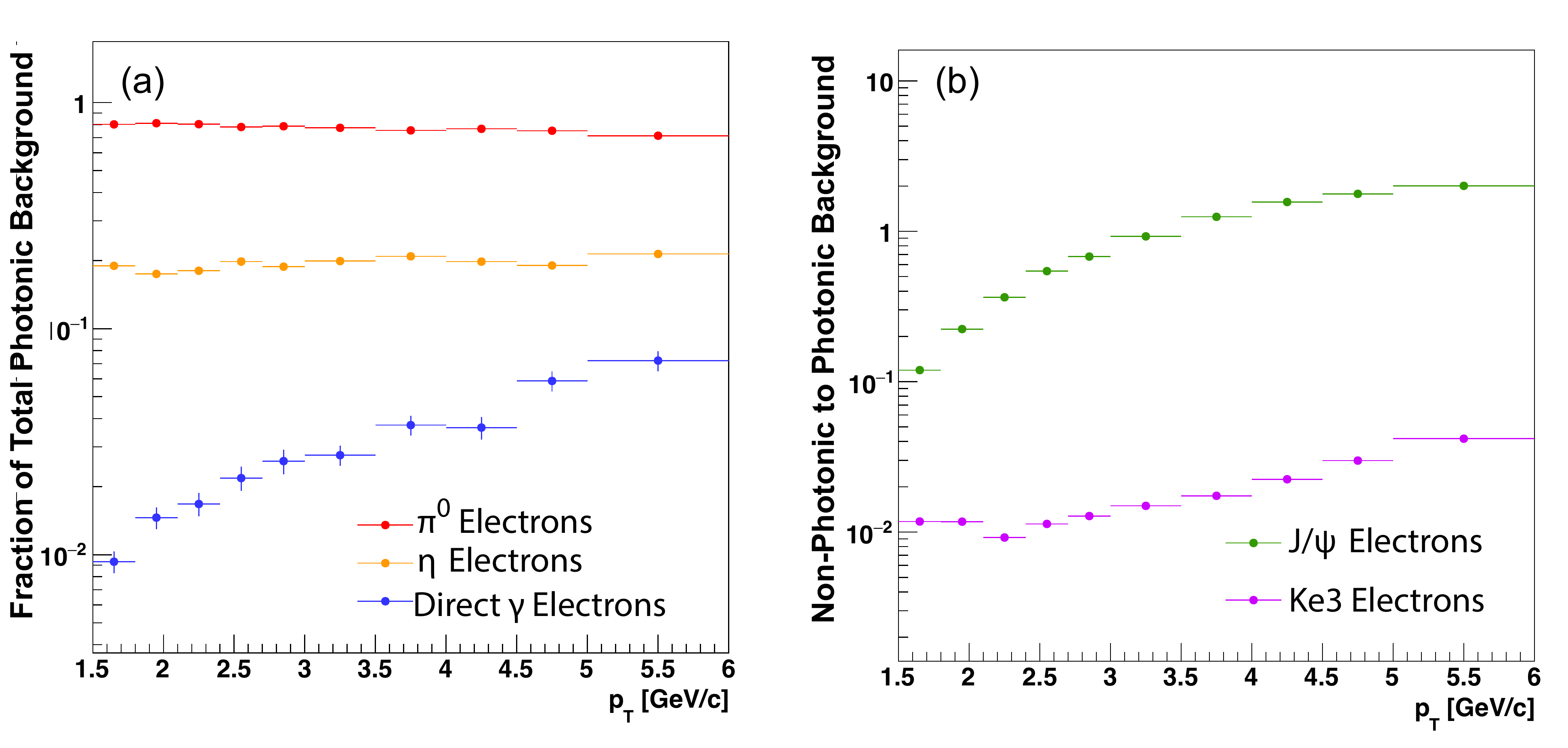}
\caption{(a) Fraction of electrons from individual photonic sources 
relative to the total photonic electron background as determined by 
constructing a background electron cocktail. (b) Fraction of electrons 
from individual nonphotonic sources relative to the total photonic 
background.}
\label{fig_cocktail}
\end{figure*}

The approach to modeling nonphotonic background is similar to that used 
for photonic sources. Namely, single particles are generated according 
to their respective published cross section, as measured by the PHENIX 
collaboration~\cite{Adare:2011vq,Adare:2011vy,Adare:2010fe}, forced to 
decay, and the resulting particles fed through a \textsc{geant3} 
simulation of the detector. Applying the full set of analysis track 
cuts, including the conversion veto cut, we complete the background 
electron cocktail. The fraction of nonphotonic electrons from each 
source, relative to the total photonic background is shown in 
Fig.~\ref{fig_cocktail}(b). Notice that the $J/\psi$ contribute more to 
the background cocktail than any other background source above 
$p_T\approx 3.5$ GeV/$c$.

\subsubsection{Hadron Contamination}

Despite the electron identification cuts described in the previous 
section, some hadron tracks will incorrectly be tagged as electrons. 
This contribution to the electron sample is modest and is estimated in 
two independent ways, making use of EMCal and RICH signals.

Unlike hadrons, electron tracks deposit the majority of their energy in 
the EMCal, as quantified by the ratio $E/p$, where $E$ is the 
calorimeter energy and $p$ is the track momentum. The variable $dep$, as 
previously defined, takes the shape of a Gaussian of zero mean and unit 
width for true electron tracks. In contrast, the $dep$ distribution of 
hadron tracks exhibits a very different shape. Thus, a template is 
constructed, as a function of $p_T$, by fitting the $dep$ distribution 
of hadrons tracks in data. The $dep$ distribution of electron candidates 
is then fit with a combination of the hadron template plus a Gaussian, 
with a single free parameter corresponding to their relative 
contribution. The value of this parameter provides an estimate of the 
fraction of hadron contamination in the sample.

An independent way of estimating the hadron contamination is to exploit 
the fact that imposing a cut requiring a minimum number of PMTs fired in 
the RICH provides greater rejection power for hadron tracks than for 
electrons. The fraction of hadrons rejected by such a cut can be 
estimated from hadron tracks in data, while the fraction of rejected 
electrons can be determined through \textsc{geant3} simulation of single 
electrons. With these two pieces of information it is possible to 
isolate the number of hadrons and electrons in the candidate electron 
sample, thus determining the contamination fraction.

%%%%%%%%%%%%%%%%%%%%%%%%%%%%%%%%%%%%%%%%%%%%%%%%%%  Fig_8
\begin{figure}[htb]
\includegraphics[width=1.0\linewidth]{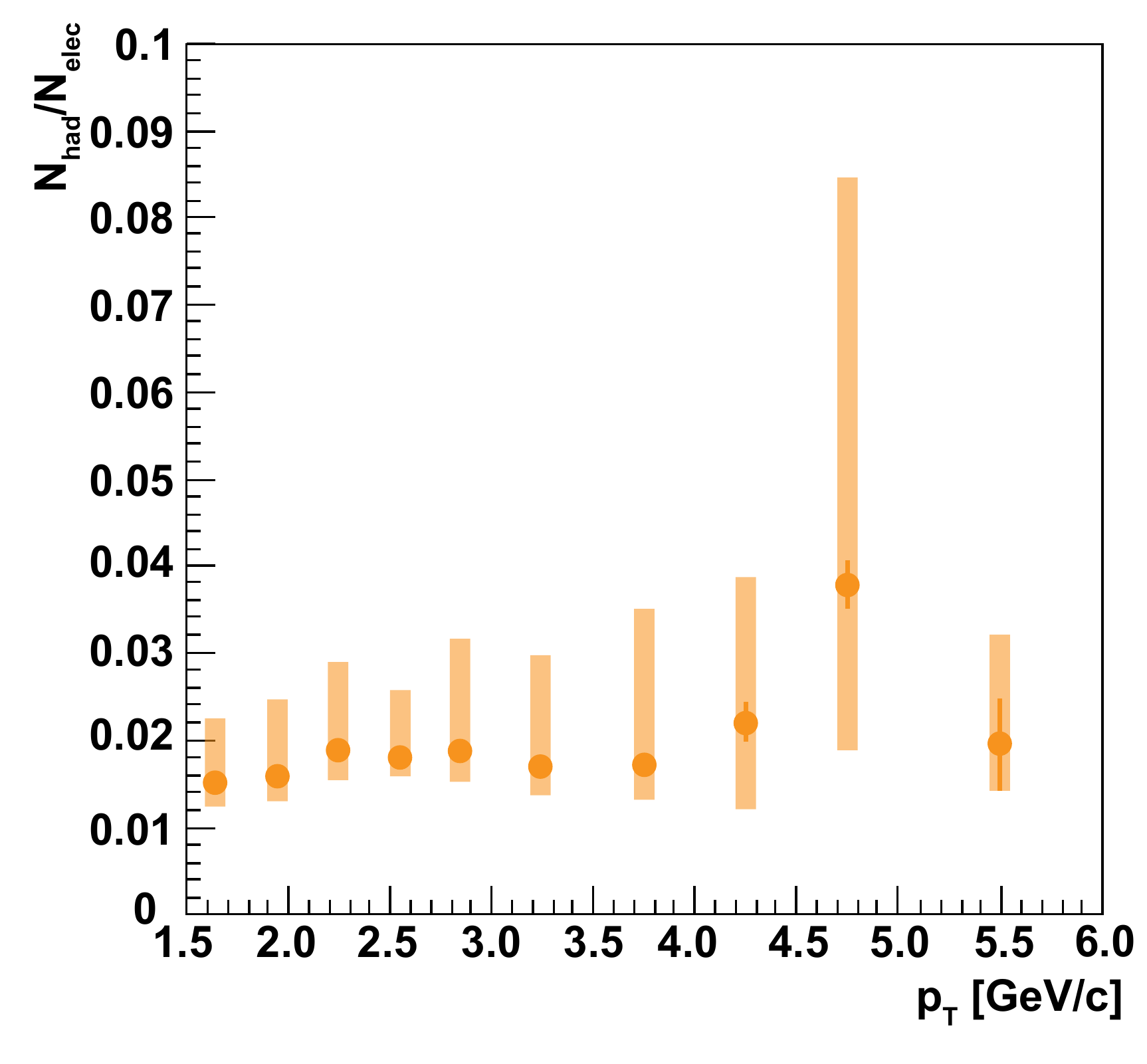}
\caption{Fraction of electron candidates in data attributed to hadron 
contamination as a function of $p_T$.}
\label{fig_frac_hadcontam}
\end{figure}

The weighted average of the two independent estimates of hadron 
contamination is taken as the nominal value, with their difference as a 
systematic uncertainty, as shown in Fig.~\ref{fig_frac_hadcontam}. The 
systematic uncertainties are assigned to encompass both estimates and 
are asymmetric.

Another source of contamination comprises electron tracks identified in 
the central arms which are associated with uncorrelated random hits in 
the VTX detector, leading to the creation of a spurious VTX-associated 
track. The degree of contamination arising in this manner was quantified 
by rotating all hits in the VTX in azimuth and polar angles by a small 
amount and attempting to re-associate central arm tracks with the 
rotated hits. Given the low multiplicity of $p$$+$$p$ collisions, the 
contribution of misassociated central-arm tracks was found to be 
negligible, unlike in Au$+$Au collisions where it is significant.

\subsection{Normalizing Electron Background DCA$_T$}
\label{subsec_normaliz}

In 2015---the year in which the $p$$+$$p$ data was collected---the VTX 
detector exhibited a time-varying acceptance from a changing number of 
dead, cold, and hot channels across the surface of each detector layer 
over time. This precluded the measurement of an electron candidate 
sample fully corrected for acceptance and efficiency effects. As a 
result, the simulated electrons in the background cocktail are not 
corrected for acceptance and efficiency, but simply constructed in such 
a way that the same reconstruction code and analysis cuts used in data 
are applied.

The cocktail can then be used to calculate the fraction of electrons 
from each background source relative to the total photonic background. 
However, to use this information to determine their normalization 
relative to the total sample of electron candidates, it is necessary to 
determine the fraction of electron candidates attributable to photonic 
background. This is accomplished via a data-driven method relying on the 
conversion veto cut.

%%%%%%%%%%%%%%%%%%%%%%%%%%%%%%%%%%%%%%%%%%%%%%%%%%  Fig_9
\begin{figure}[htb]
\includegraphics[width=1.0\linewidth]{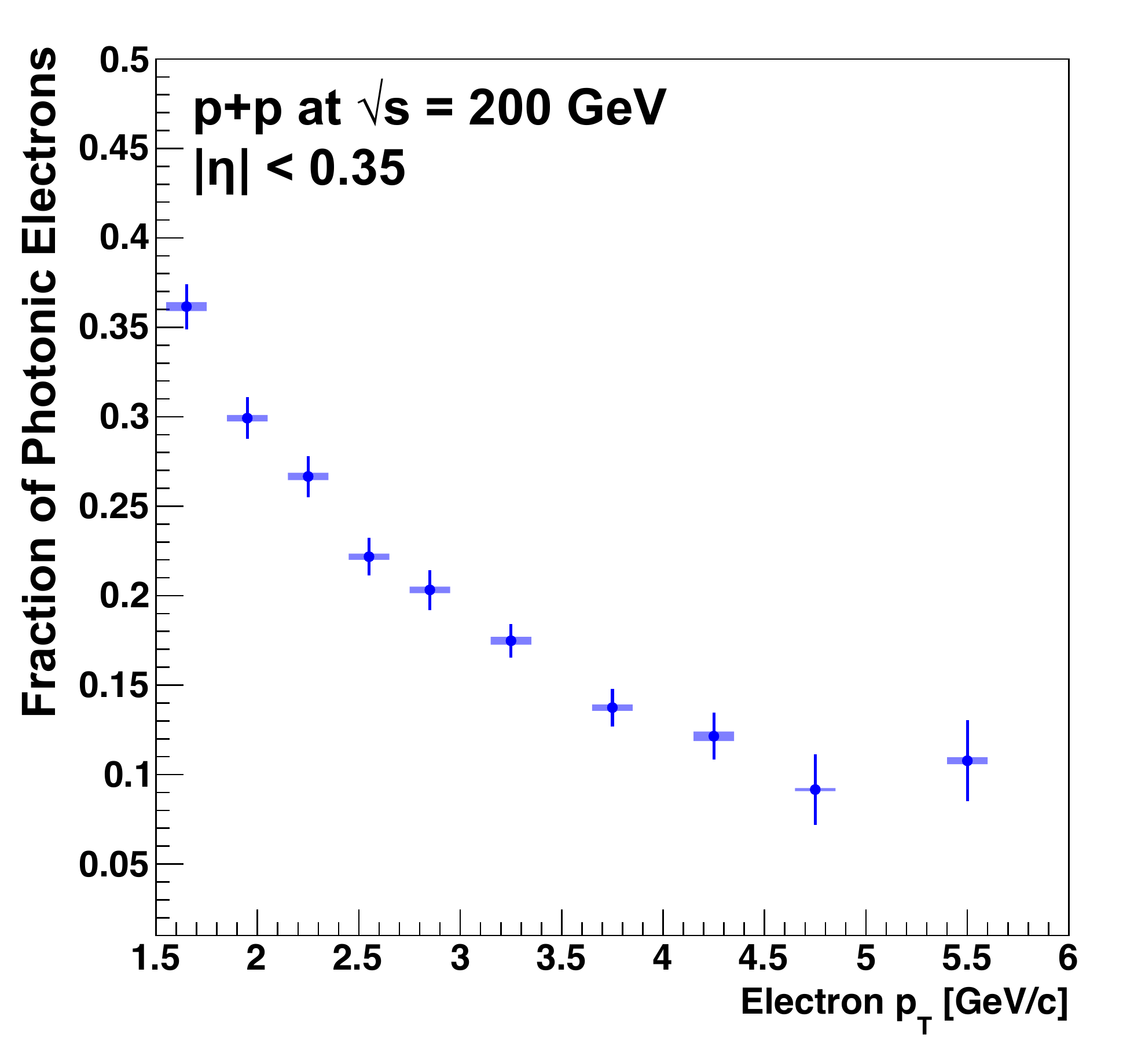}
\caption{Fraction of photonic electrons to inclusive electrons in data, 
as a function of electron $p_T$.}
\label{fig_fp}
\end{figure}

Let $N_P$ and $N_{\rm NP}$ be the number of photonic and nonphotonic 
electrons in the electron candidate sample obtained without applying the 
conversion veto cut. Also, let $\varepsilon_P$ and $\varepsilon_{UC}$ be 
the veto cut survival rate of photonic electrons due to correlated 
effects and due to noncorrelated effects, respectively. In this 
nomenclature, heavy-flavor electrons are part of the nonphotonic sample. 
The number of electrons measured without the conversion veto cut is then 
simply
\begin{equation}
N_e = N_P + N_{\rm NP},
\label{eq_ne}
\end{equation}
while the number of measured electrons that pass the veto cut is given by
\begin{equation}
\tilde{N}_e = \varepsilon_P \times \varepsilon_{UC} \times N_{P} + \varepsilon_{UC}\times N_{\rm NP},
\label{eq_ne_tilde}
\end{equation}
where $N_{P}$ is modified by both $\varepsilon_P$ and 
$\varepsilon_{UC}$ because photonic electrons are also susceptible to 
rejection from uncorrelated hits in the window. Taken together, 
Eqs.~\ref{eq_ne} and ~\ref{eq_ne_tilde} form a system of equations with 
$N_P$ and $N_{\rm NP}$ as the only unknowns, yielding
\begin{equation}
N_{P} = \frac{\tilde{N}_e - 
N_e\varepsilon_{UC}}{\varepsilon_{UC}(\varepsilon_{P}-1)},
\end{equation}
and
\begin{equation}
N_{\rm NP} = \frac{N_e\varepsilon_{P}\varepsilon_{UC} - 
\tilde{N}_e}{\varepsilon_{UC}(\varepsilon_{P}-1)}.
\end{equation}
The fraction of photonic electrons in the sample with the conversion 
veto cut applied is then
\begin{equation}
F_{P} = \frac{\varepsilon_P\varepsilon_{UC} 
N_{P}}{\varepsilon_P\varepsilon_{UC} N_{P} + \varepsilon_{UC} N_{\rm NP}},
\end{equation}
and is shown in Fig.~\ref{fig_fp} as a function of electron $p_T$. Using 
$F_{P}$, the fraction of candidate electrons attributable to each 
photonic source in data is
\begin{equation}
f_i^{{\rm phot}} = F_P(1-F_{{\rm 
contam}})\frac{\tilde{N}_i}{\tilde{N}_{\pi^0} + \tilde{N}_{\eta} + 
\tilde{N}_{\gamma}},
\end{equation}
where $i$ is an index referring to a primary particle species (i.e., 
$\pi^0, \eta, \gamma$); $\tilde{N}_i$ is the number of electrons from 
the $i^{{\rm th}}$ source that pass the conversion veto cut in the 
electron cocktail, and $1-F_{{\rm contam}}$ is the purity of the 
electron sample from hadron contamination.

In the case of nonphotonic background, it is impossible to construct a 
similar expression because the electron cocktail does not include 
contributions from heavy-flavor mesons. Therefore, the nonphotonic 
background is normalized relative to the simulated $\pi^0$ electron 
yield, whose absolute normalization has been previously determined, as 
follows
\begin{equation}
f_j^{{\rm nonphot}} = f_{\pi^0}^{{\rm phot}}
\frac{\tilde{N}_j}{\tilde{N}_{\pi^0}},
\end{equation} 
with $j$ indexing the primary particles giving rise to nonphotonic 
electrons (i.e., $J/\psi$ and $Ke3$).

%%%%%%%%%%%%%%%%%%%%%%%%%%%%%%%%%%%%%%%%%%%%%%%%%%  Fig_10
\begin{figure}[htb]
\includegraphics[width=1.0\linewidth]{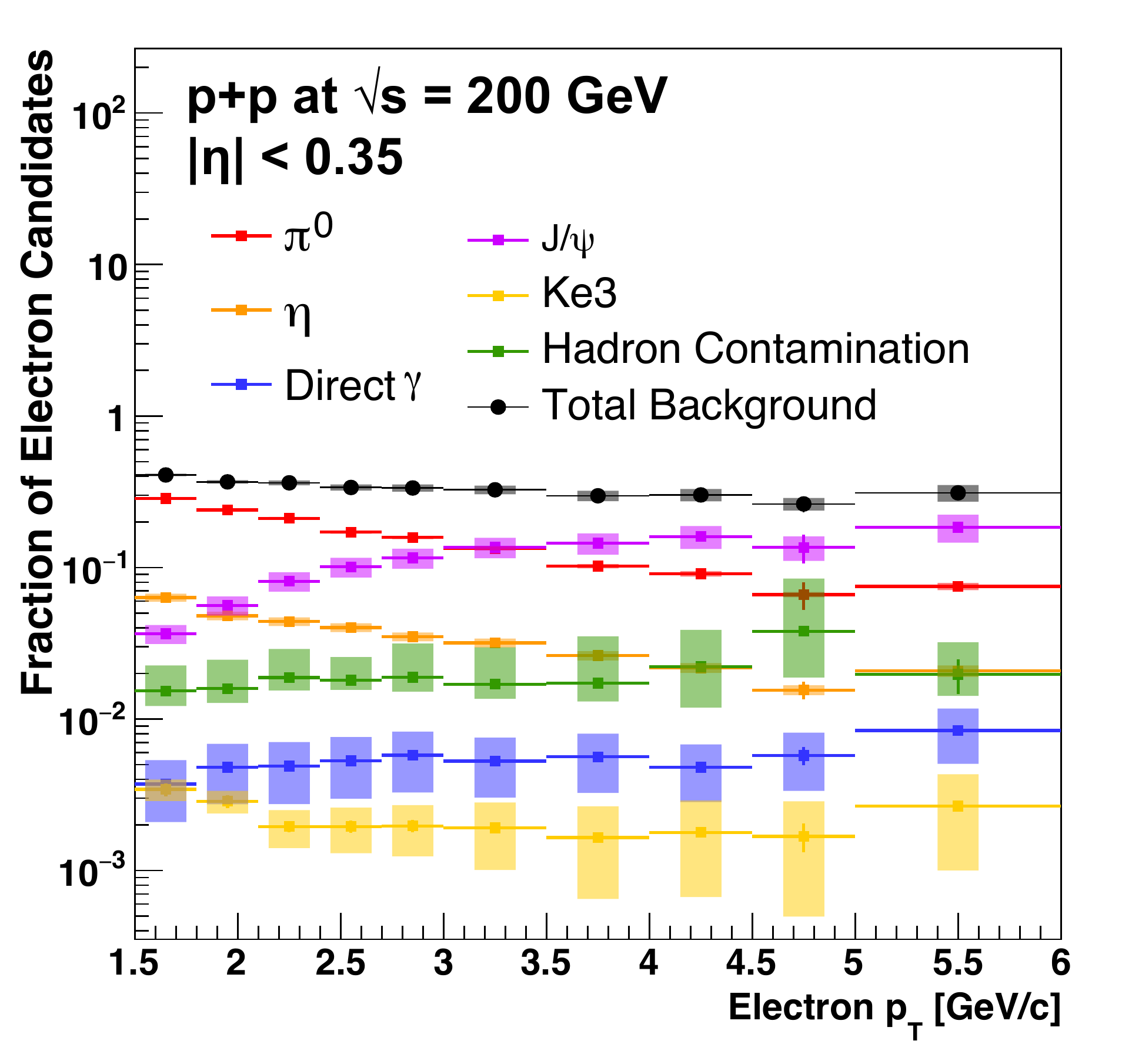}
\caption{Fraction of the measured candidate electron sample attributable 
to various sources of background electrons as a function of $p_T$.}
\label{fig_cocktail_fractions}
\end{figure}

%%%%%%%%%%%%%%%%%%%%%%%%%%%%%%%%%%%%%%%%%%%%%%%%%%  Fig_11
\begin{figure}[htb]
\includegraphics[width=1.0\linewidth]{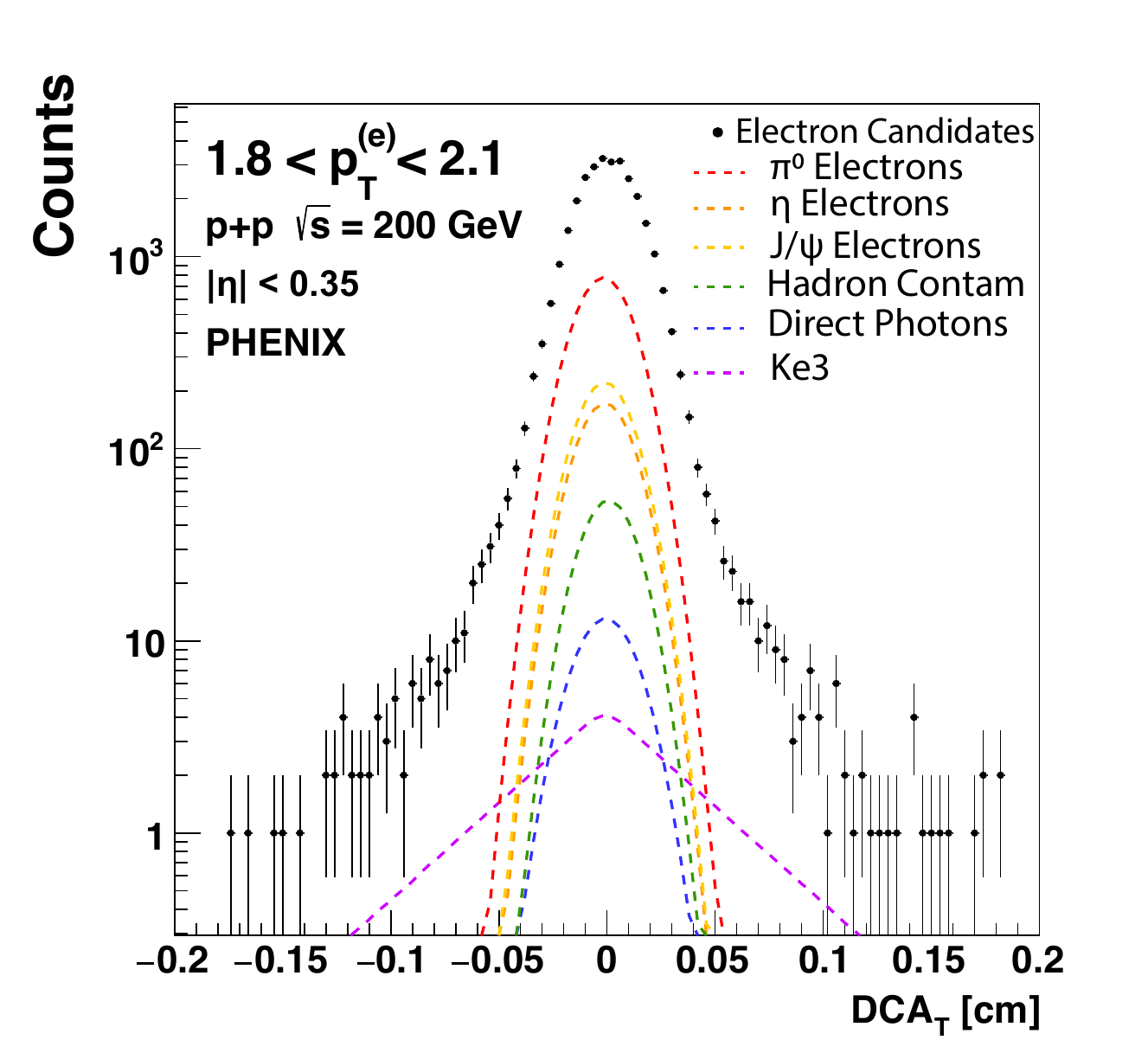}
\caption{DCA$_T$ distribution of candidate electron tracks within $1.8 < 
p_T{\rm [GeV/}c] < 2.1$, which pass the conversion veto cut. Also shown 
are the absolutely normalized DCA$_T$ distributions from each simulated 
background electron source, as well as hadron contamination.}
\label{fig_dca_electron}
\end{figure}

These factors, shown in Fig.~\ref{fig_cocktail_fractions}, are used to 
normalize the DCA$_T$ distribution of each background electron source 
relative to the electron candidate sample. Fig.~\ref{fig_dca_electron} 
shows the DCA$_T$ distribution of electrons from each background species 
within $1.8 < p_T{\rm [GeV/}c] < 2.1$, normalized relative to the total 
number of electron candidates in that $p_T$ bin. Unlike prompt 
electrons, which exhibit a Gaussian DCA$_T$ shape, K$e$3 electrons 
originate from long-lived kaon decays, which results in their 
characteristic DCA$_T$ shape.

%======================================================  Table_III
\begin{table}
\caption{Resolution of the DCA$_T$ distribution of charged pion tracks 
as reconstructed in data and in Monte Carlo \textsc{geant3} simulations 
of the PHENIX detector.}
\begin{ruledtabular} \begin{tabular}{ccccc}
& Electron $p_T$ & Data Resolution & MC Resolution &  \\
&    [GeV/$c$]   &     [$\mu$m]    &     [$\mu$m]  &  \\
\hline
& $1.5 < p_T < 1.8$ & 134.1 & 122.1 & \\
& $1.8 < p_T < 2.1$ & 131.7 & 121.0 & \\
& $2.1 < p_T < 2.4$ & 130.1 & 119.9 & \\
& $2.4 < p_T < 2.7$ & 128.9 & 118.9 & \\
& $2.7 < p_T < 3.0$ & 127.9 & 118.5 & \\
& $3.0 < p_T < 3.5$ & 126.8 & 118.2 & \\
& $3.5 < p_T < 4.0$ & 126.2 & 117.9 & \\
& $4.0 < p_T < 4.5$ & 125.5 & 117.6 & \\
& $4.5 < p_T < 5.0$ & 125.1 & 117.2 & \\
& $5.0 < p_T < 6.0$ & 124.8 & 116.9 & \\
\end{tabular} \end{ruledtabular}
\label{table_dca_res}
\end{table}

In general, the resolution of the DCA$_T$ is a consequence of the width 
of the beam spot convolved with the track-pointing resolution. However, 
the simulations used to create the background electron cocktail were run 
using a single reference value for the beam spot size, which in reality 
fluctuates over time during data-taking. Therefore, it is necessary to 
correct for the difference in resolution between simulations and data. 
This was accomplished by comparing, as a function of $p_T$, the DCA$_T$ 
resolution of charged pions in data and simulation, as quoted in 
Table~\ref{table_dca_res} deriving a $p_T$-dependent factor such that 
the simulated DCA$_T$ distributions could be broadened to match the 
resolution in data. This broadening factor was applied to electrons from 
every species in the background electron cocktail.

\subsection{Heavy-Flavor Separation via Unfolding}
\label{subsec_unfold}

Having normalized the electron background DCA$_T$ distributions, it is 
possible to isolate the corresponding distributions of electrons from 
heavy-flavor decays in data. If the shapes of the parent hadron spectra 
were known a priori, it would be a straightforward matter to use the 
knowledge of heavy-flavor-decay kinematics to determine the shape of the 
DCA$_T$ distributions of charm and bottom electrons separately, whose 
relative normalization could then be constrained by the measured 
inclusive DCA$_T$.

%%%%%%%%%%%%%%%%%%%%%%%%%%%%%%%%%%%%%%%%%%%%%%%%%%  Fig_12
\begin{figure}[htb]
\includegraphics[width=1.0\linewidth]{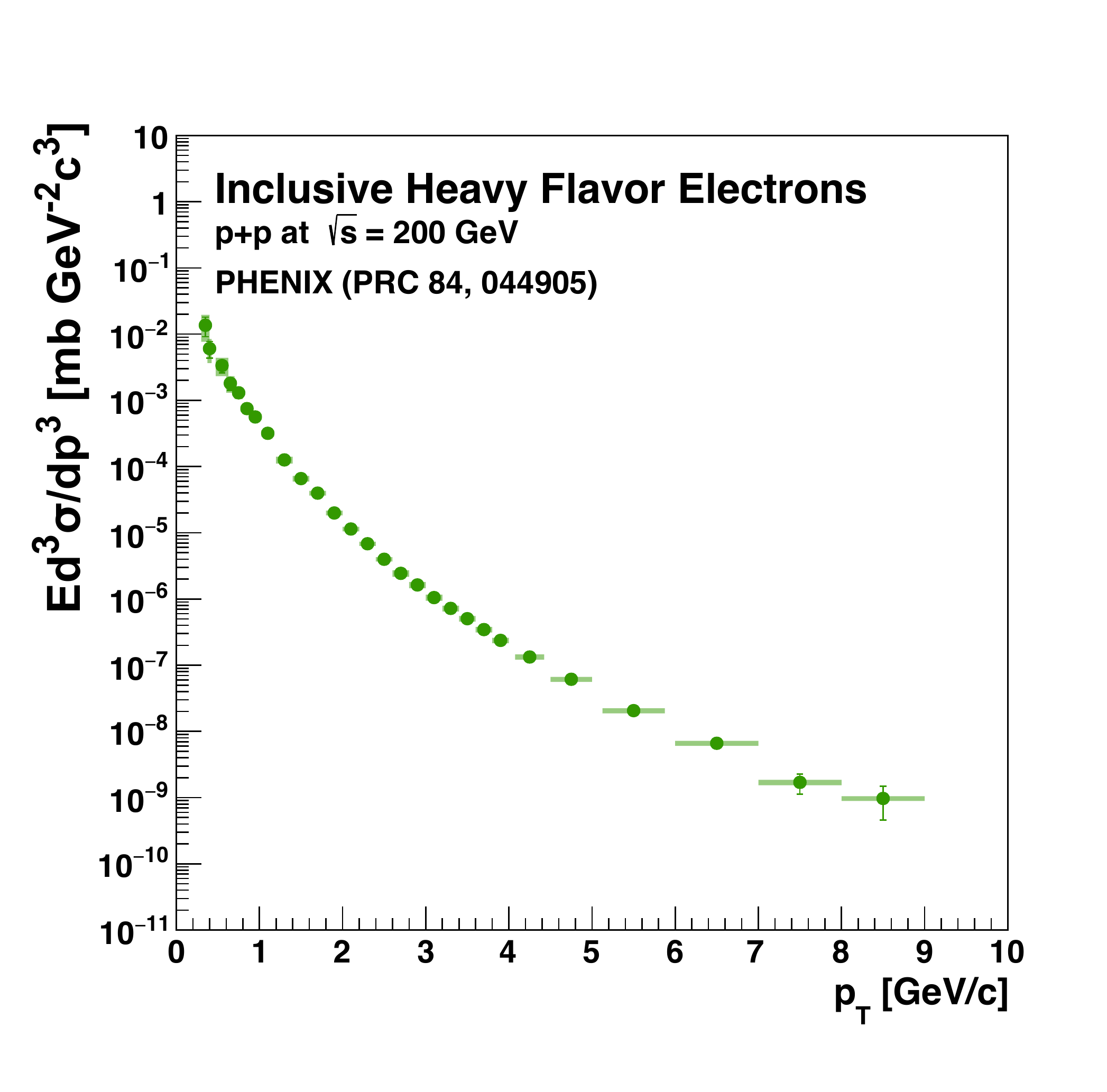}
\caption{Invariant cross section of inclusive heavy-flavor-electron 
production in $p$$+$$p$ collisions, as measured by the PHENIX 
experiment~\cite{Adare:2010de}, and used as input for the 
flavor-electron-separation-unfolding procedure.}
\label{fig_hfe_yield}
\end{figure}

However, because such spectral shapes are not known for $D$ and $B$ 
mesons, it becomes necessary to solve an inverse problem where the model 
parameters (i.e., the spectrum of hadrons containing open charm and 
bottom, as a function of $p_T$) are inferred from data observations, 
namely the DCA$_T$ and spectrum of inclusive heavy-flavor electrons. For 
the spectrum, an earlier PHENIX measurement~\cite{Adare:2010de} was 
used, shown in Fig.~\ref{fig_hfe_yield}.

To solve the inverse problem, it is necessary to construct a mapping 
from the model parameters to the data.  Given that the 
heavy-flavor-decay kinematics are known, it is possible to assign a 
probability for a heavy flavor hadron at a given $p_T^{(h)}$ to decay 
into an electron with a certain $p_T^{(e)}$ and DCA$_T$. Such mapping 
makes it possible to quantify the likelihood that a given set of trial 
hadron spectra is consistent with the measured electron spectrum and 
DCA$_T$ distributions.

We use a probabilistic approach~\cite{Choudalakis:2012hz} to the 
unfolding problem based on Bayesian inference, identical to that used by 
PHENIX to separate charm and bottom electron yields in Au$+$Au 
collisions~\cite{Adare:2015hla}. Let $\mathbf{Y}^{{\rm data}}$ be a 
vector whose individual elements correspond to the yield of inclusive 
heavy-flavor electrons, as shown in Fig.~\ref{fig_hfe_yield}. Similarly, 
$\mathbf{D}_j^{{\rm data}}$ is a vector of the binned DCA$_T$ 
distribution of electrons in data, for tracks in the $j^{{\rm th}}$ 
$p_T$ bin, out of nine bins between $1.5 < p_T {\rm [GeV}/c] < 6.0$. The 
two observables are combined in a `data' vector

\begin{equation}
\bm{x} = (\mathbf{Y}^{{\rm data}}, \mathbf{D}_0^{{\rm data}},\ldots,\mathbf{D}_8^{{\rm data}}).
\end{equation}

The model parameters are also represented as a vector 
\begin{equation}
\bm{\theta}=(\bm{\theta}_c,\bm{\theta}_b), 
\end{equation}
where $\bm{\theta}_{c}$ and $\bm{\theta}_{b}$ correspond to the charm 
and bottom hadron yields, respectively, in 17 $p_T$ bins each, between 
$0 < p_T{\rm [GeV}/c] < 20$.

Bayes' theorem, as written below
\begin{equation}
P(\bm{\theta}\mid \bm{x}) = \frac{P(\bm{x}\mid 
\bm{\theta})\pi(\bm{\theta})}{P(\bm{x})},
\end{equation}
relates the probability that a given set of model parameters 
$\bm{\theta}$ are true given the data $\bm{x}$, to the probability that 
the data follow from an assumed set of model parameters. While the 
former probability---known as the \textit{posterior}---is a difficult 
quantity to estimate, the latter---known as the \textit{likelihood}---is 
straightforward to compute given the knowledge of heavy-flavor decays. 
The quantity $\pi(\bm{\theta})$, known as the \textit{prior}, 
corresponds to the knowledge of the model parameters prior to the data 
being analyzed. The denominator $P(\bm{x})$, sometimes known as the 
\textit{evidence}, provides the normalization for the posterior. Thus, 
Bayes' theorem allows us to take a first guess regarding the model 
parameters, as encoded in the prior, and refine it through the inclusion 
of data in the likelihood.

The 17 bins for both the charm and bottom hadron spectra within $0 < 
p_T^{(h)}{\rm [GeV}/c] < 20$, as represented by $\bm{\theta}$, define a 
34-dimensional space of model parameters. Starting with an initial set 
of values given by the prior $\pi(\bm{\theta})$, corresponding to the 
charm and bottom yields as calculated with 
\textsc{pythia}\footnote{We used \textsc{pythia6.2}, with 
parton distribution functions (PDFs) 
given by CTEQ5L. The following parameters were 
modified: MSEL = 5, MSTP(91) = 1, PARP(91) = 1.5, MSTP(33) = 1, PARP(31) 
= 2.5. For bottom (charm) hadron studies, PARJ(13) = 0.75 (0.63), 
PARJ(2) = 0.29 (0.2), PARJ(1) = 0.35 (0.15).}, the unfolding proceeds by 
drawing trial sets of hadron yields, corresponding to individual points 
in the multi-dimensional parameter space. Because sampling such a 
large-dimensional space uniformly is computationally prohibitive, we use 
a Markov chain Monte Carlo (MCMC) 
algorithm~\cite{1538-3873-125-925-306}, which proceeds iteratively until 
convergence of the final solution is achieved. In this analysis, three 
iterations suffice for convergence with 500 parallel ``walkers'' and 
1000 burn-in steps, as described in~\cite{1538-3873-125-925-306}.

For each trial $\bm{\theta}$, we predict an electron $p_T$ spectrum and 
DCA$_T$ distribution as follows

\begin{equation}
\bm{Y}(\bm{\theta}) = \bm{M}^{(\bm{Y})}\bm{\theta}_c + 
\bm{M}^{(\bm{Y})}\bm{\theta}_b,
\end{equation}
\begin{equation}
\bm{D}_j(\bm{\theta}) = \bm{M}_j^{(\bm{D})}\bm{\theta}_c + 
\bm{M}_j^{(\bm{D})}\bm{\theta}_b,
\end{equation}
where $\bm{M}^{\bm{Y}}:p_T^{(h)}\rightarrow p_T^{(e)}$ is a matrix 
encoding the probability of a hadron of $p_T^{(h)}$ of any rapidity to 
yield an electron of $p_T^{(e)}$ at midrapidity, while 
$\bm{M}_j^{(\bm{D})}:p_T^{(h)}\rightarrow {\rm DCA}_T^{(e)}$ encodes the 
probability of yielding an electron at a given DCA$_T$ value. The 
construction of these matrices using the \textsc{pythia} generator is 
described in detail in Ref.~\cite{Adare:2015hla}, and includes the 
decays of charm hadrons ($D^{\pm}$,$D^0$,$D_s$, and $\Lambda_c$), and 
bottom hadrons ($B^{\pm}$, $B^0$, $B_s$m and $\Lambda_b$). Additional 
Monte Carlo generators could be used to construct the matrix, but this 
would be computationally prohibitive. For the purposes of this analysis, 
an additional matrix was introduced to model the detector response, 
mapping the \textit{truth} $p_T$ and DCA$_T$ values in 
$\bm{M}^{(\bm{Y})}$ and $\bm{M}_j^{(\bm{D})}$ to their 
\textit{reconstructed} counterparts, allowing for a direct comparison 
between the data and the predicted distributions from a given set of 
trial parameters $\bm{\theta}$.

The predicted spectrum and DCA$_T$ distributions are then used to 
compute the (log)likelihood as follows

\begin{equation}
\begin{split}
\ln P(\bm{x}\mid\bm{\theta}) &=\ln P(\bm{Y}^{{\rm 
data}}\mid\bm{Y}(\bm{\theta}))\\ &+ \sum_{j=0}^{8}\ln P(\bm{D}_j^{{\rm 
data}}\mid\bm{D}_j(\bm{\theta})),
\end{split}
\end{equation}
where the log-likelihood for the $\bm{Y}^{{\rm data}}$ term is modeled 
as a multi-variate Gaussian with diagonal covariance, while the 
log-likelihood for the $\bm{D}_j$ term is modeled by a multivariate 
Poisson distribution, with full details provided in 
Ref.~\cite{Adare:2015hla}.

To constrain the shape of the unfolded charm and bottom spectra, 
ensuring its smoothness, a regularization term is added to the 
log-likelihood function, as follows

\begin{equation}
\ln \pi(\bm{\theta}) = -\alpha^2(|\bm{LR}_c|^2+|\bm{LR}_b|^2),
\end{equation}
where $\bm{R}_c$ and $\bm{R}_b$ correspond to the ratios of the trial 
vector of charm and bottom spectra to the prior. The matrix $\bm{L}$ is 
a $17\times 17$ discretized second-order finite-difference matrix, 
effectively corresponding to the second derivative operator. Thus, the 
addition of this term enhances the log-likelihood for solutions with 
large curvature, effectively penalizing deviations from smoothness. The 
optimal regularization strength $\alpha$ is determined by carrying out a 
scan of various possible parameter values and calculating the total 
log-likelihood of the unfolding solution in each case, comparing it to 
the case with no regularization. The desired optimal value is that which 
maximizes the log-likelihood, and is found to be $\alpha=1$.

%%%%%%%%%%%%%%%%%%%%%%%%%%%%%%%%%%%%%%%%%%%%%%%%%%  Fig_13
\begin{figure*}[tbh]
\includegraphics[width=0.99\linewidth]{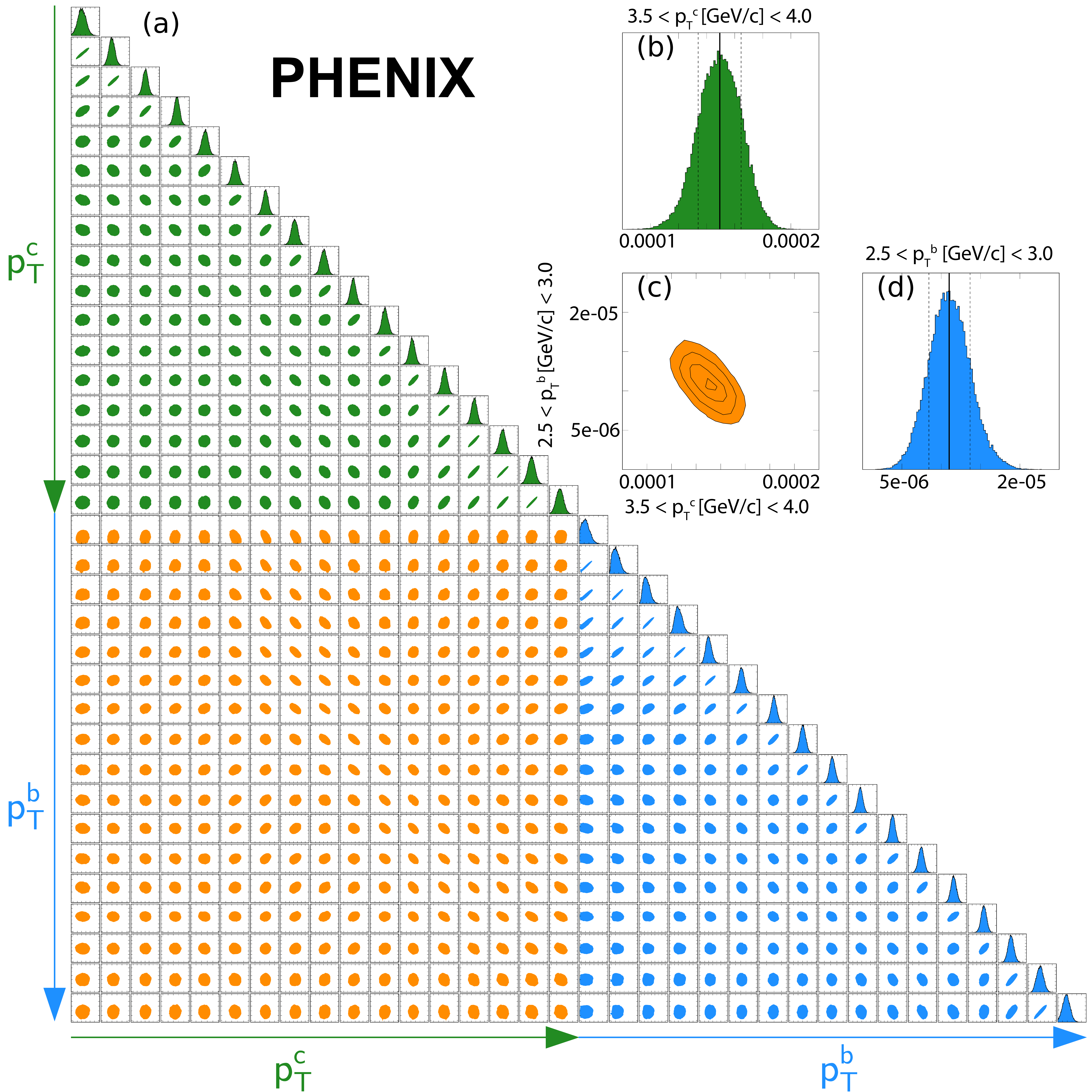}
\caption{(a) Joint probability distributions of bottom and charm hadron 
yields. The marginalized distribution for each $p_T$ bin is shown along 
the diagonal.  Correlations among bins are shown in 
the upper triangular [green] area for $p_T^c$ and $p_T^c$, 
the far-right triangular [blue] area for $p_T^b$ and $p_T^b$, 
and the lower-left [orange] area for $p_T^b$ and $p_T^c$. 
(b)-(d) Correlation 
between charm and bottom yields in two particular $p_T$ bins, along with 
the marginalized distributions in those bins. See text for details.} 
\label{fig_triangle} 
\end{figure*}

The end result of the Monte Carlo exploration of the parameter space is 
a set of probability distributions for each of the 34 model parameters, 
corresponding to the value of each bin of the charm and bottom hadron 
spectra integrated over all rapidities, including the correlations among 
them, as depicted in Fig.~\ref{fig_triangle}(a). The diagonal of the 
triangle shows the marginal probability distribution for each of the 17 
bins of the charm and bottom hadron spectra.
Correlations among bins are shown in
the upper triangular [green] area for $p_T^c$ and $p_T^c$,
the far-right triangular [blue] area for $p_T^b$ and $p_T^b$,
and the lower-left square [orange] area for $p_T^b$ and $p_T^c$.
Panels Fig.~\ref{fig_triangle}(b) and (d) show the 
marginal distributions in detail for charm and bottom hadrons in two 
selected $p_T$ bins. We select the parameter that maximizes the marginal 
distributions as the desired value of the spectrum at each bin; the 
16$^{{\rm th}}$ and 84$^{{\rm th}}$ quantiles of the distribution are 
taken as the 1$\sigma$ uncertainty associated with the point estimate, 
as indicated by the dotted lines.

Panel Fig.~\ref{fig_triangle}(c) shows the joint probability 
distribution of the charm and bottom hadron yields for the $p_T$ bins in 
Fig.~\ref{fig_triangle}(b) and (d). The shape of the distribution 
indicates the existence of a strong negative correlation between the 
yields in the bins at hand.  It is possible to see that the bins are 
largely uncorrelated, except for intermediate $p_T^b$ and $p_T^c$, where 
a strong negative correlation exists due to the yields being similar in 
this kinematic region.  For the correlations among bins in the same 
hadron spectrum, a very strong positive correlation is seen among 
neighboring $p_T$ bins owing to the smoothness requirement on the 
unfolded spectra, which is imposed via regularization.

%%%%%%%%%%%%%%%%%%%%%%%%%%%%%%%%%%%%%%%%%%%%%%%%%%  Fig_14
\begin{figure*}[tbh]
\includegraphics[width=0.99\linewidth]{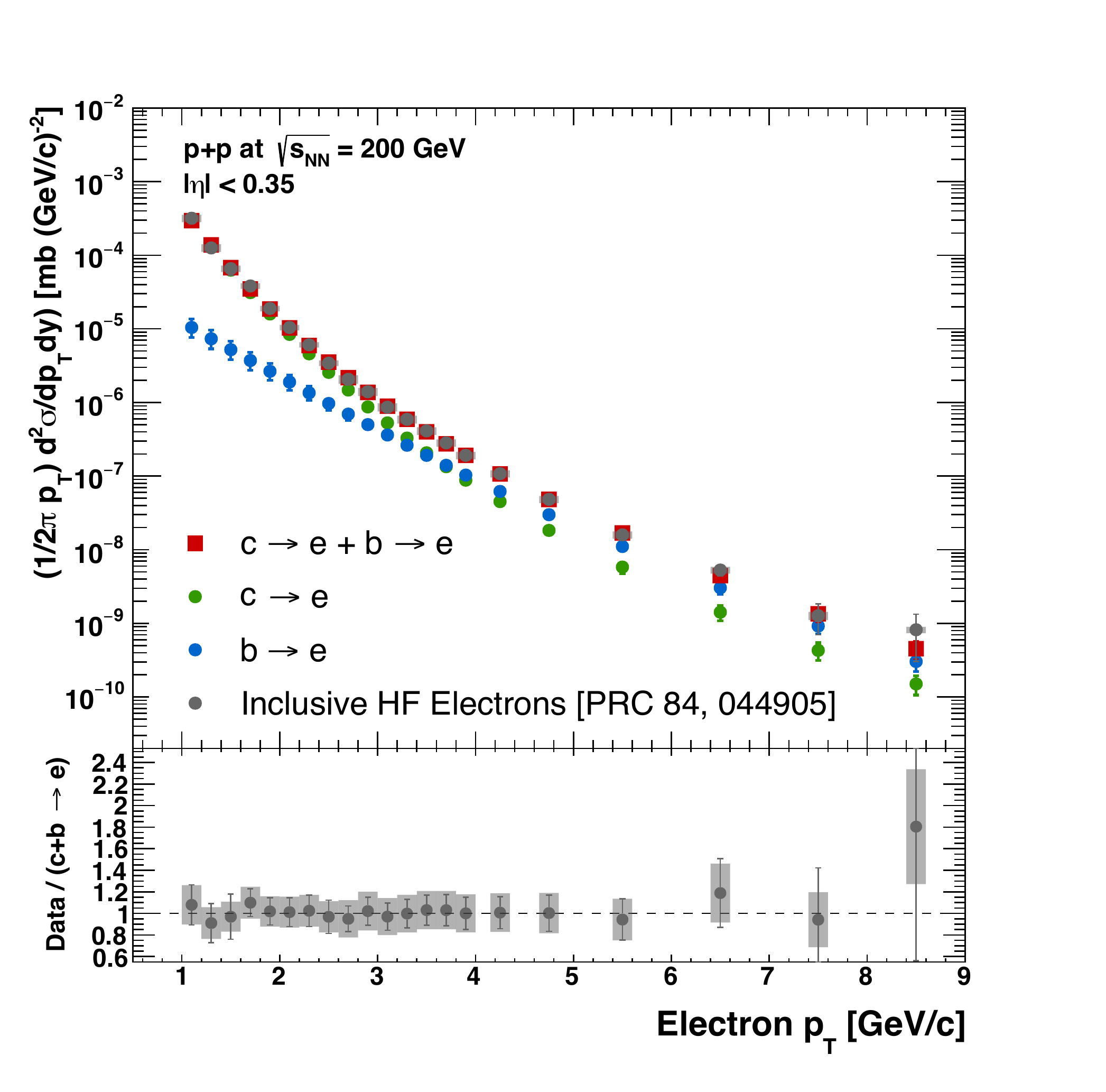}
\caption{Inclusive heavy-flavor-electron invariant yield from the 
refolded charm and bottom yields (closed squares [red]) compared to 
published data (closed circles [gray]). Individual refolded spectra 
from charm and bottom are shown in green and blue, respectively.}
\label{fig_ept_refold}
\end{figure*}

%%%%%%%%%%%%%%%%%%%%%%%%%%%%%%%%%%%%%%%%%%%%%%%%%%  Fig_15
\begin{figure*}[htb]
\includegraphics[width=0.99\linewidth]{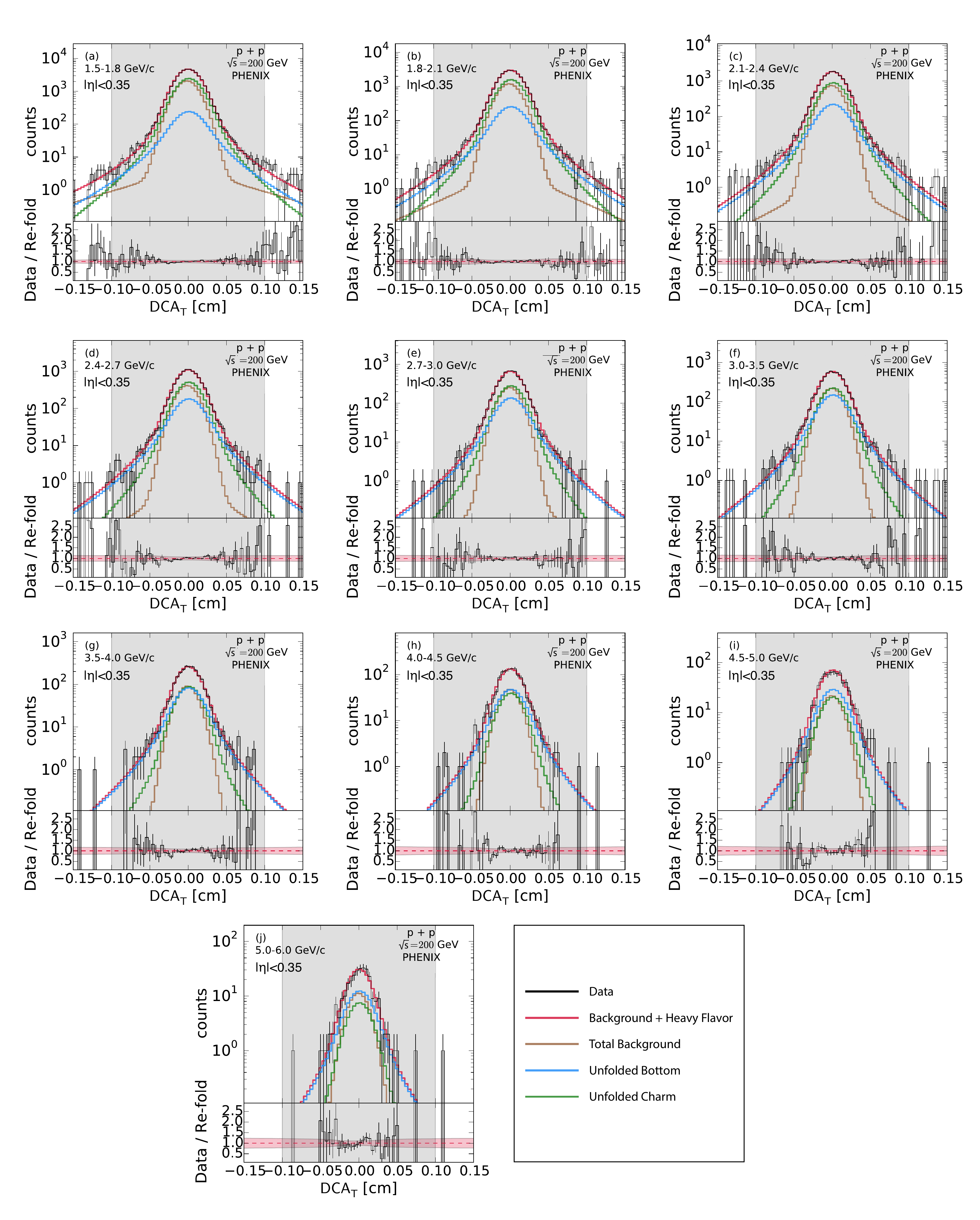}
\caption{DCA$_T$ distribution of electron candidates in various $p_T$ 
bins, along with the contribution from total background electrons 
(brown) and the refolded electrons from charm (green) and bottom (blue) 
hadron decays. The sum of these three components is shown in red, and 
the ratio with the data is shown in the bottom panels. The shaded gray areas
indicate the region over which the DCA$_T$ provides constraints for the
unfolding procedure.}
\label{fig_dca_refold}
\end{figure*}

The unfolded yield of charm and bottom hadrons can be tested for 
consistency with the inputs provided, namely the spectrum and DCA$_T$ 
distributions of inclusive heavy-flavor electrons, by applying the decay 
matrices to the unfolded result.  Figure~\ref{fig_ept_refold} shows the 
so-called `refolded' spectra of charm (bottom) electrons in green 
(blue), along with their sum, in red.  The refolded inclusive spectrum 
compares very well with the published spectrum, as shown by the ratio 
plot in the bottom panel.  Similarly, Fig.~\ref{fig_dca_refold} shows, 
for every electron $p_T$ bin, the refolded inclusive electron DCA$_T$ 
distributions, its charm and bottom components, and the total background 
DCA$_T$, obtained as discussed in section~\ref{subsec_normaliz}. The 
ratio plots in the bottom panel demonstrate an excellent agreement with 
the DCA$_T$ of measured electrons. The shaded gray region indicates the 
range over which the DCA$_T$ is used in the unfolding procedure. Notice 
that bottom electrons have a broader DCA$_T$ than those from charm, as 
expected.

\section{Systematic Uncertainties}
\label{sec_systematics}

The unfolding procedure, as described in section~\ref{subsec_unfold}, 
takes the measured electron DCA$_T$ distributions and published electron 
spectrum as inputs, along with their corresponding statistical 
uncertainties, which are propagated to the final result. However, 
additional sources of systematic uncertainty must be taken into account. 
Namely, we identify the following as the most significant: 

\begin{enumerate}

\item the normalization of individual sources in the background electron 
cocktail;

\item the systematic uncertainty on the inclusive heavy-flavor electron 
spectrum; 

\item the choice of the regularization parameter strength $\alpha$ in 
the unfolding procedure; and

\item the choice of prior used in the unfolding.

\end{enumerate}

The uncertainty associated with the normalization of individual electron 
background components originates from the parameterization of the 
associated primary particle spectrum. Each spectrum is repeatedly 
deformed randomly within the extent of its own statistical and 
systematic uncertainties, with a new parameterization being obtained at 
every iteration. The RMS value of all parameterizations is then taken as 
the associated systematic uncertainty. In this manner, a systematic 
uncertainty will exist for every background electron source in the 
cocktail. Their combined effect on the unfolded result is estimated by 
running the unfolding procedure for every combination of individual 
background normalizations, raised and lowered by their associated 
parameterization uncertainty.

To estimate the $p_T$-correlated systematic uncertainties associated 
with the inclusive heavy-flavor-electron spectrum, we deform the shape 
of the spectrum by tilting and kinking the curve about two pivot points, 
at $p_T=2.5$ GeV/$c$ and $p_T=5.0$ GeV/$c$. The choice of these points 
is motivated by specific features of the previous analysis which 
produced the inclusive heavy-flavor-electron 
spectrum~\cite{Adare:2010de}, related to the method of background 
subtraction. Tilting refers to a rotation of the spectrum about one of 
the two pivots, such that the first and last points go up and down, 
respectively, by a fraction of their systematic uncertainty. The kinking 
of the spectrum introduces a deformation whereby the spectrum takes on a 
``v'' shape at the pivots. This procedure resulted in 8 variations of 
the spectrum. The ones that resulted in the largest deviation from the 
nominal unfolded result were taken as the associated systematic 
uncertainty.

Section~\ref{subsec_unfold} described how the optimal value of the 
regularization strength $\alpha$ maximizes the total log-likelihood of 
the unfolded solution. An uncertainty on this value is determined by 
finding the values of $\alpha$ around the maximum which lead to a 
decrease of the log-likelihood by half a unit, effectively corresponding 
to a 1$\sigma$ uncertainty. The deviations of the unfold result obtained 
with these values ($\alpha = 0.71$ and $\alpha = 1.55$), relative to the 
\textit{nominal} result when using the optimal parameter, define the 
extent of the associated systematic uncertainty.

Finally, a systematic uncertainty is associated with the choice of 
$\bm{\theta}_{{\rm prior}}$. The magnitude of this uncertainty is 
estimated by selecting a different prior and evaluating the change in 
the unfold result. In particular, the heavy-flavor-hadron yields 
obtained with \textsc{pythia} were scaled by a modified blast wave 
calculation, as described in Ref.~\cite{Adare:2013wca}. Because a 
feature of Markov chains, such as the one used in this analysis, is that 
the probability of reaching a given state is independent of the starting 
point, the sensitivity to the initial choice of prior is expected to be 
minimal after a sufficient number of iterations.

%%%%%%%%%%%%%%%%%%%%%%%%%%%%%%%%%%%%%%%%%%%%%%%%%%  Fig_16
\begin{figure}[htb]
\includegraphics[width=1.0\linewidth]{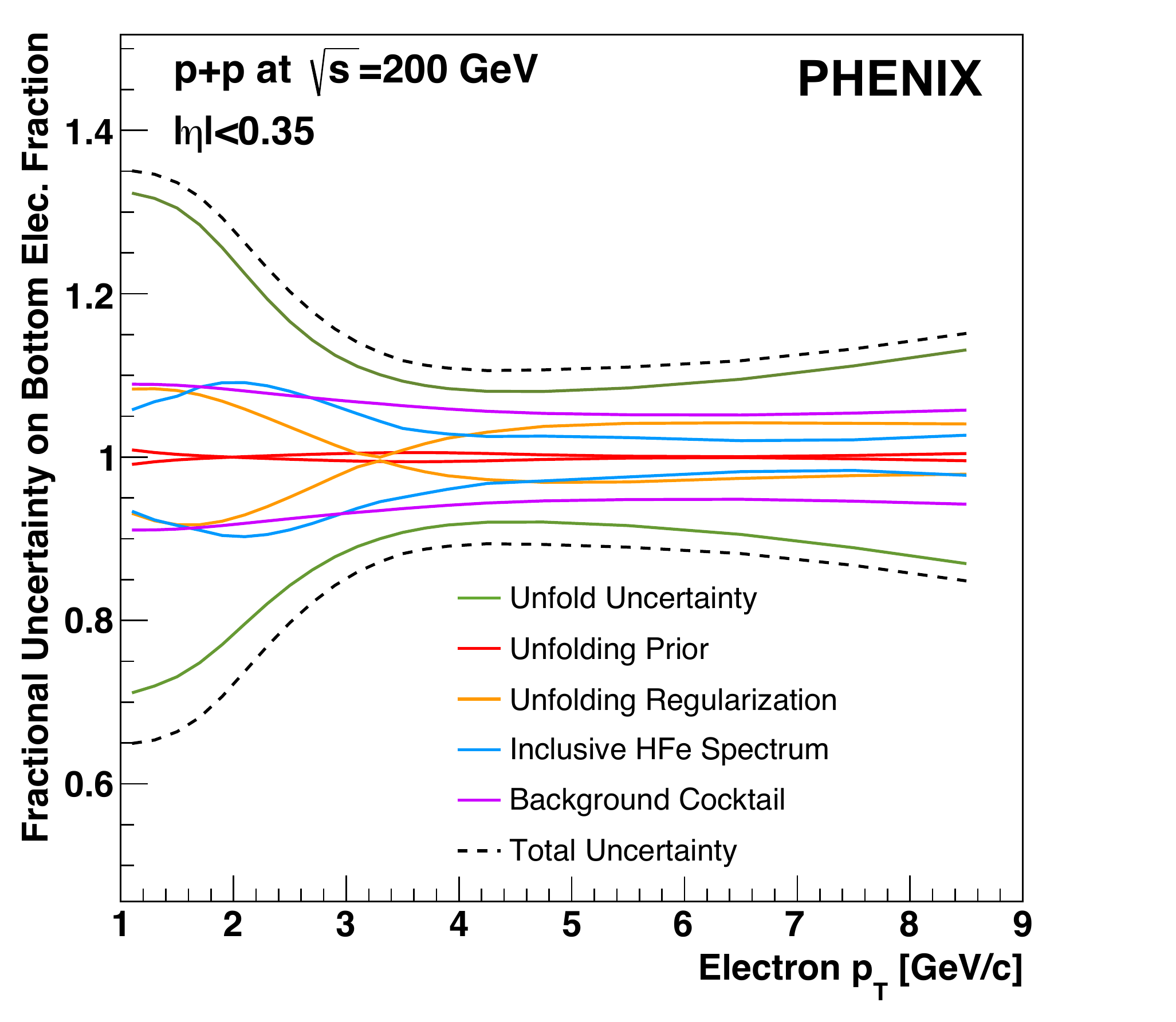}
\caption{Fractional uncertainty on the bottom electron fraction, defined 
as $1\pm$ the relative uncertainty of each source.}
\label{fig_frac_syst}
\end{figure}

Figure~\ref{fig_frac_syst} shows the relative contribution of each source 
of uncertainty as a function of $p_T$, to the unfolded fraction of 
electrons from bottom decays. The most significant contribution comes 
from the unfold uncertainty, which originates from the statistical 
uncertainty on the inclusive heavy-flavor spectrum and DCA$_T$ as it is 
propagated through the unfold procedure. The next most significant 
contribution comes from the background electron cocktail and its 
normalization, supplying an approximate 10\% uncertainty at low $p_T$. 
The total systematic uncertainty is obtained by adding the contributions 
of every source in quadrature.

\section{Results and Discussion}

\begin{figure}[htb]
\includegraphics[width=1.0\linewidth]{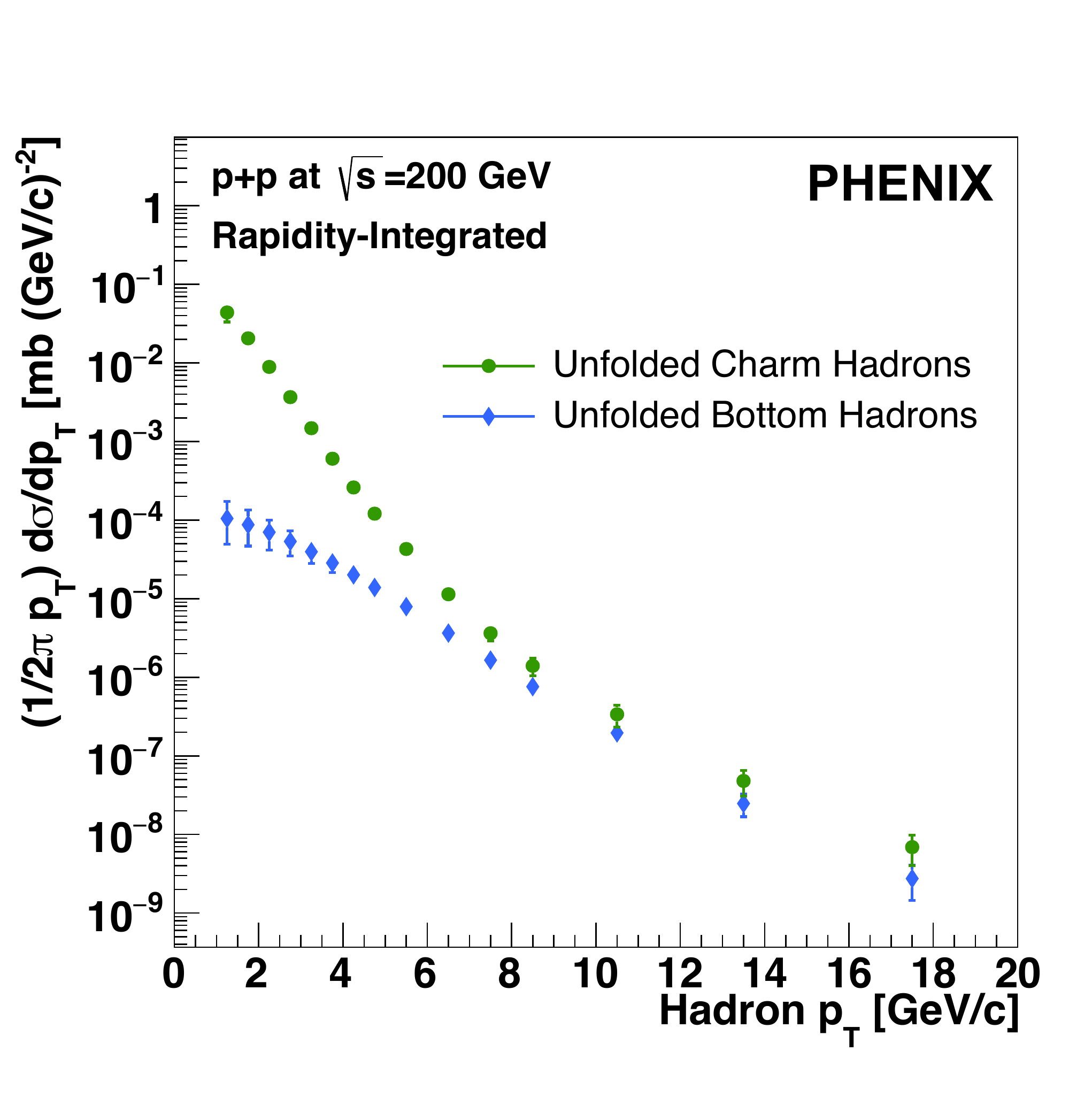}
\caption{Unfolded charm and bottom hadron yields in bins of of 
transverse momentum. The yields are integrated over all rapidities, and 
the error bars depict the combined statistical and systematic 
uncertainties.}
\label{fig_cb_yield}
\end{figure}

Figure~\ref{fig_cb_yield} shows the invariant cross section of charm and 
bottom hadrons integrated over all rapidity, corresponding to the values 
that maximize the probability distribution associated with each hadron 
$p_T$ bin, shown along the diagonal of Fig.~\ref{fig_triangle}. The 
uncertainties incorporate both the unfolding uncertainty (which includes 
the propagation of the total uncertainty in the inclusive 
heavy-flavor-electron measurements provided as input) as well as the 
systematic uncertainties discussed in section~\ref{sec_systematics}. The 
uncertainty band is narrowest in the region where electron DCA$_T$ 
measurements provide constraint to the unfold procedure, namely $1.5 < 
p_T {\rm [GeV/}c] < 6.0$. As presented, the hadron cross section is 
integrated over rapidity by construction, following directly from the 
procedure used to populate the decay matrices used in the unfolding 
procedure. Namely, hadrons simulated in \textsc{pythia} at all 
rapidities are allowed to decay, recording only the probability of 
producing an electron within $|\eta|<0.35$. It thus follows that the 
cross sections in Fig.~\ref{fig_cb_yield} depend on the hadron rapidity 
distribution implemented in the \textsc{pythia} generator. This model 
dependence implies an associated uncertainty which has not been 
evaluated since, as previously mentioned, this would be computationally 
prohibitive. Furthermore, the model dependence is reduced when applying 
the decay matrix to arrive at results in electron space.

%%%%%%%%%%%%%%%%%%%%%%%%%%%%%%%%%%%%%%%%%%%%%%%%%%  Fig_17

\begin{figure}[htb]
\includegraphics[width=1.0\linewidth]{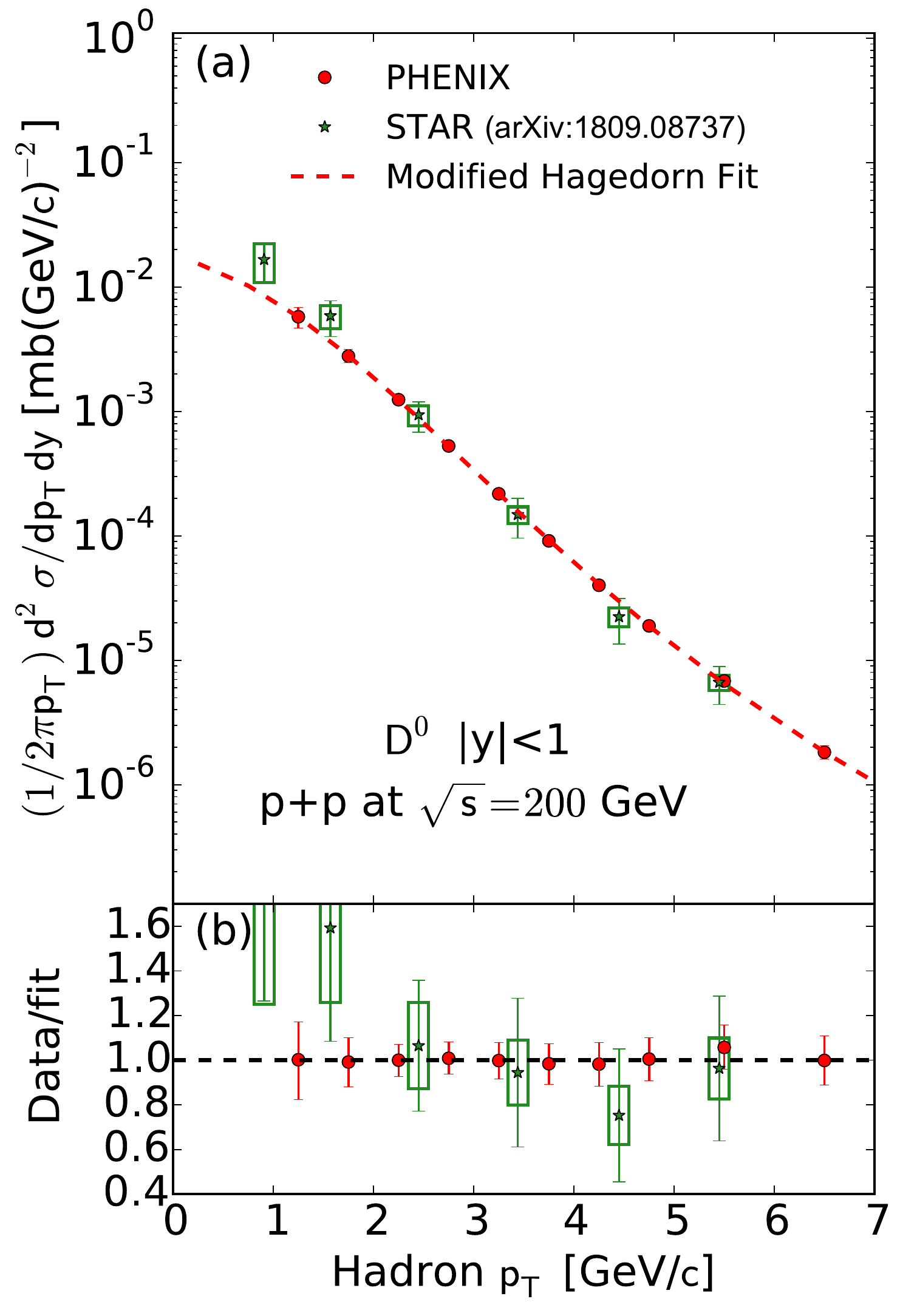}
\caption{Unfolded differential cross section of $D^0$ mesons at 
midrapidity $|y|<1$, compared to a corresponding measurement by the STAR 
experiment~\cite{Adamczyk:2014uip} obtained by direct reconstruction of 
the hadron decays.}
\label{fig_star_d0}
\end{figure}

To compare the unfolded differential hadron cross sections to existing 
measurements, we use the \textsc{pythia} generator to calculate the 
ratio of open heavy-flavor hadrons of a given species at midrapidity 
relative to inclusive-hadron production as a function of $p_T$. In this 
manner, the unfolded yield of $D^0$ mesons within $|y|<1$ can be 
compared to a measurement by the STAR 
collaboration~\cite{Adamczyk:2012af} obtained by fully reconstructing 
the hadron decays, as shown in Fig.~\ref{fig_star_d0}. The unfolded 
$D^0$ yield is fit with a modified Hagedorn function, with the ratio of 
data relative to the fit being shown in the bottom panel. Good agreement 
with results published by STAR is observed within uncertainties.

%%%%%%%%%%%%%%%%%%%%%%%%%%%%%%%%%%%%%%%%%%%%%%%%%%  Fig_18

\begin{figure*}[tbh]
\includegraphics[width=0.99\linewidth]{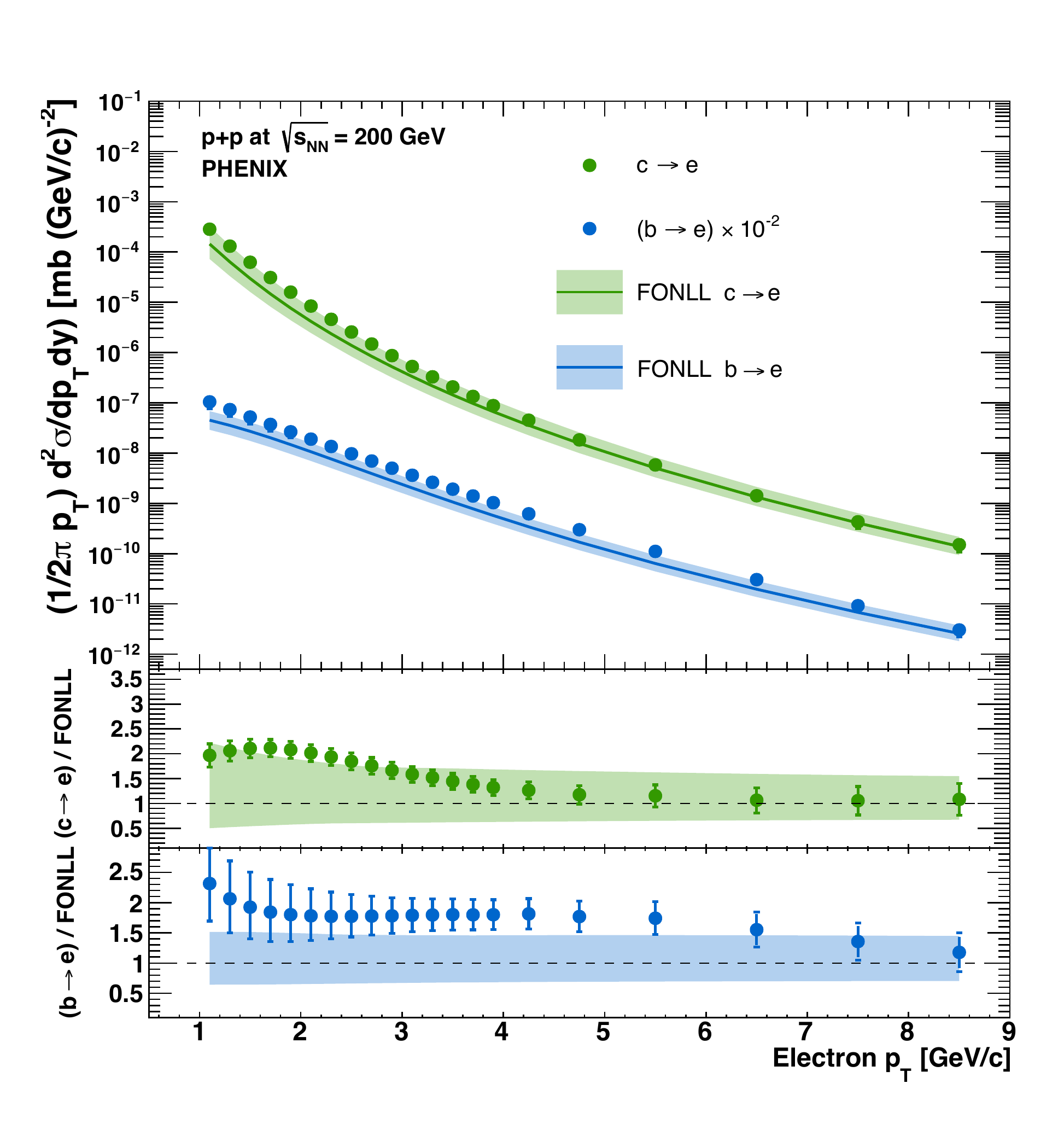}
\caption{Refolded spectra of electrons at midrapidity from charm and 
bottom decays, compared to FONLL calculations~\cite{Cacciari:2005rk}. 
The bottom electron spectrum has been scaled down by a factor of 
100 for easier comparison.}
\label{fig_refold_fonll}
\end{figure*}

The model dependence of the unfolded charm and bottom cross sections can 
be reduced by applying the decay model---that is, multiplying the hadron 
cross sections by the decay matrices---to obtain the refolded cross 
section of heavy-flavor-decay electrons in the PHENIX central arm 
acceptance. The result can be normalized to obtain the fully invariant 
differential cross sections shown in Fig.~\ref{fig_refold_fonll}, where 
the $b\rightarrow e$ curve has been scaled down by a factor of 100 for 
ease of visualization. Also shown are 
FONLL\footnote{We used the current default parameters with CTEQ6.6 PDFs.
Central values: $m_b=4.75$~GeV, $m_c=1.5$~GeV,
 $\mu_\mathrm{R}=\mu_\mathrm{F}=\mu_0=\sqrt{\mathrm{m}^2+p_T^2}$.
Scales uncertainties:
$\mu_{0}/2<\mu_\mathrm{R}$, $\mu_\mathrm{F}<2\mu_0$ with 
$1/2<\mu_\mathrm{R}/\mu_\mathrm{F}<2$.
Mass uncertainties: $m_b=4.5$, 5.0 GeV, 
$m_c=1.3$, 1.7 GeV, summed in quadrature to scales uncertainties. 
PDFs uncertainties are calculated according to the individual 
PDF set recipe, and summed in quadrature to scales and mass uncertainties. 
Branching ratios:
BR(D$\rightarrow$l)=0.103, BR(B$\rightarrow$l)=0.1086, 
BR(B$\rightarrow$D$\rightarrow$l)=0.096, BR(B$\rightarrow$D)=0.823, 
BR(B$\rightarrow$D*)=0.173, BR(B${\rightarrow}J/\psi$)=0.0116, and 
BR(B${\rightarrow}\psi(2S)$)=0.00307.}
pQCD calculations~\cite{Cacciari:2005rk}, which are in reasonable agreement 
with both charm and bottom cross sections within uncertainties. 
The large uncertainties in FONLL are driven by variations in the 
factorization and renormalization scales, as well as uncertainties in 
the heavy quark masses and the PDFs used.
The central FONLL curves in Fig.~\ref{fig_refold_fonll} correspond
to the total cross sections for charm and bottom of 
$\sigma_c(\rm FONLL)=242~\mu$b and $\sigma_b(\rm FONLL)=1.80~\mu$b.
Notice that, unlike the rapidity-integrated hadron observables in 
Figs.~\ref{fig_cb_yield} and ~\ref{fig_star_d0}, the electron spectra in 
Fig.~\ref{fig_refold_fonll} are reported at midrapidity by construction, 
following from the definition of the decay matrix. Like other 
heavy-flavor measurements at RHIC, the results presented are higher than 
the FONLL calculation. However, it is notable that the agreement with 
the central FONLL prediction improves at high $p_T$, where the effects 
of the quark mass in the calculation become less significant.

%%%%%%%%%%%%%%%%%%%%%%%%%%%%%%%%%%%%%%%%%%%%%%%%%%  Fig_19

%%%%%%%%%%%%%%%%%%%%%%%%%%%%%%%%%%%%%%%%%%%%%%%%%%  Fig_20
\begin{figure}[htb]
\includegraphics[width=1.0\linewidth]{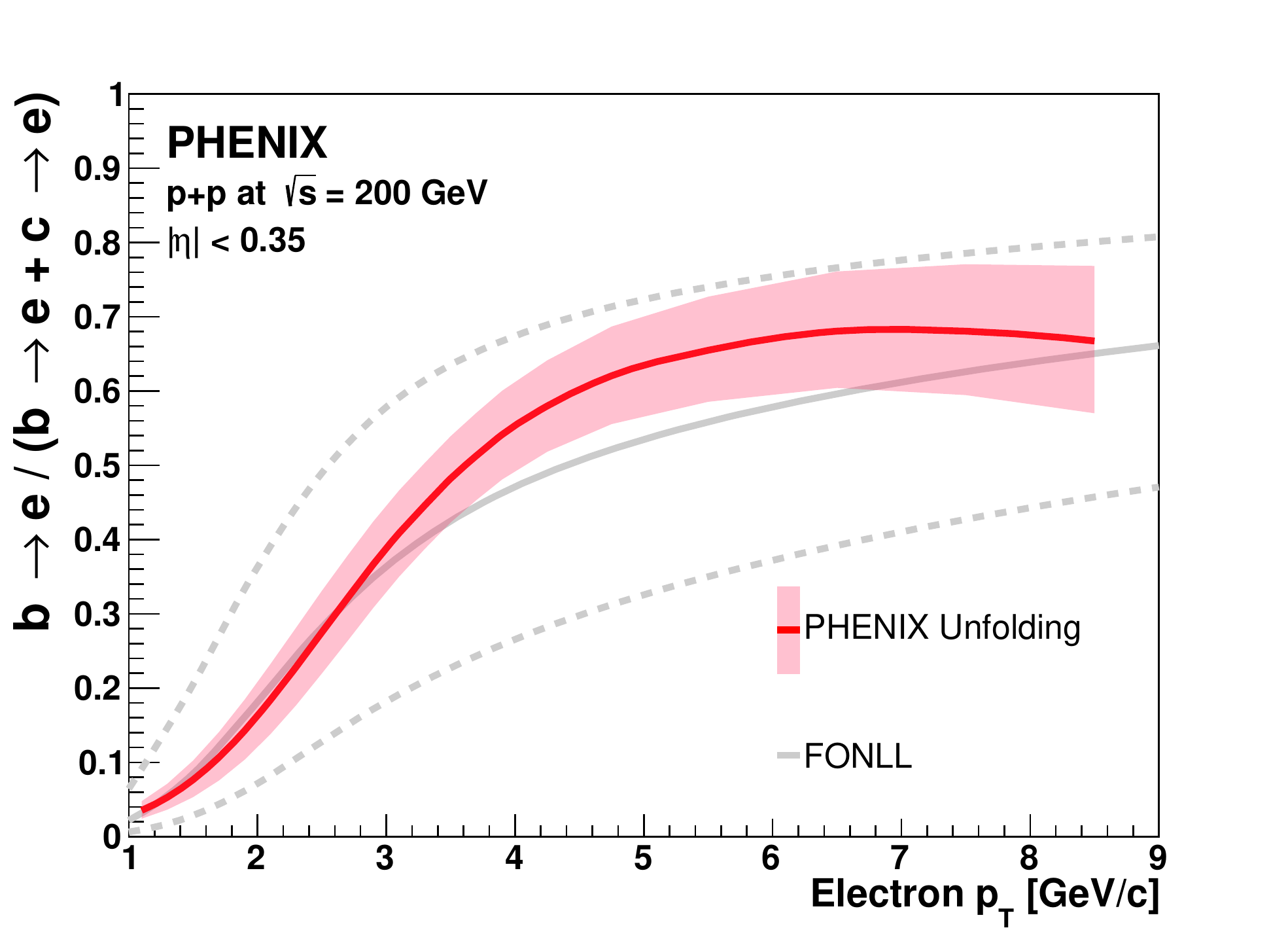}
\caption{Fraction of bottom electrons obtained with the unfolding 
procedure (red), compared to a pQCD FONLL calculation 
(gray)~\cite{Cacciari:2005rk}, with uncertainties arising from the quark 
masses and regularization scales.}
\label{fig_bfrac_fonll}
\end{figure}

%%%%%%%%%%%%%%%%%%%%%%%%%%%%%%%%%%%%%%%%%%%%%%%%%%  Fig_21
\begin{figure}[htb]
\includegraphics[width=1.0\linewidth]{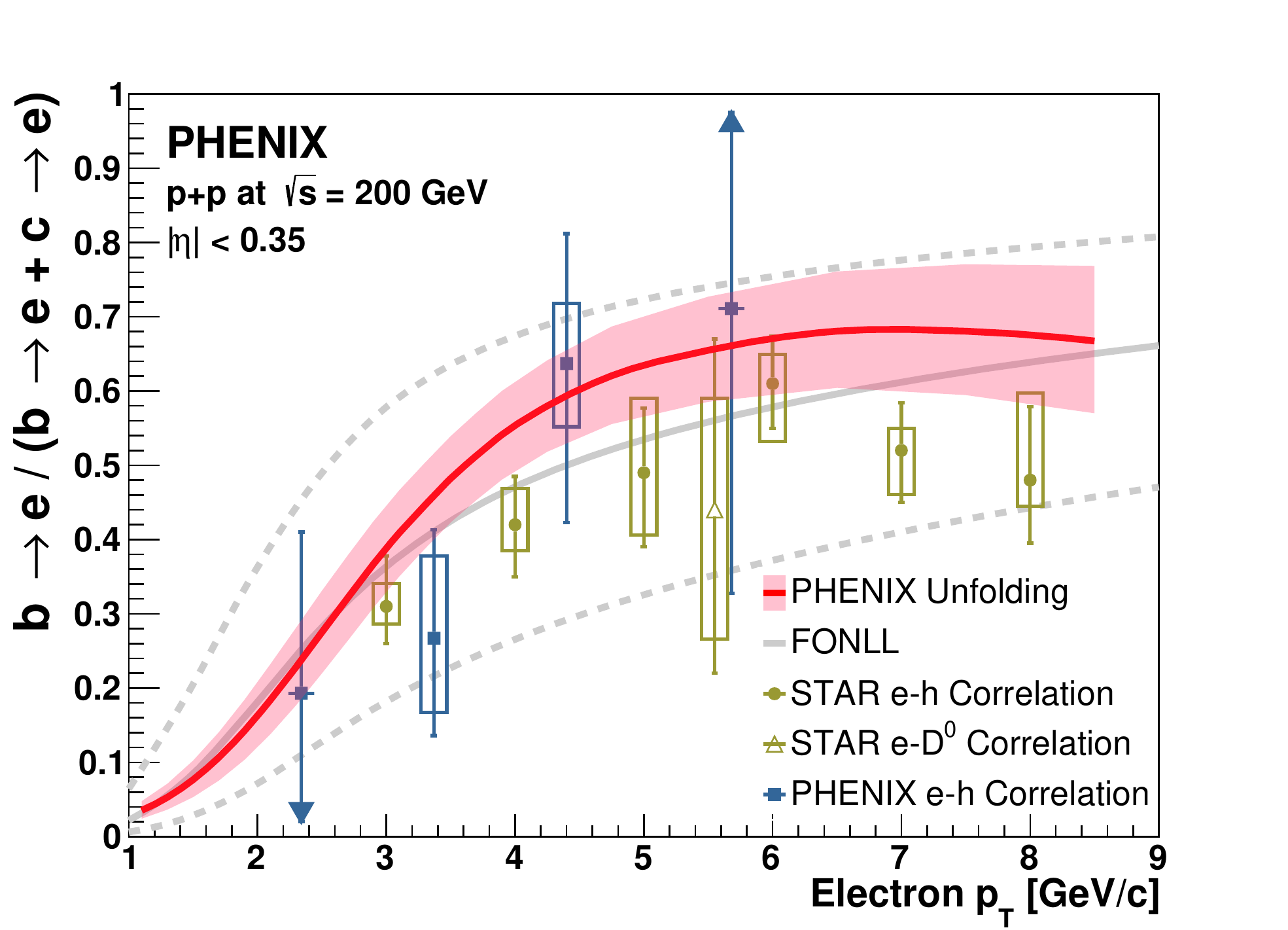}
\caption{Fraction of bottom electrons obtained with the unfolding 
procedure (red), compared to previous measurements by 
PHENIX~\cite{Adare:2009ic} and STAR~\cite{Aggarwal:2010xp} using 
electron-hadron correlations. Also shown are theory comparisons to a 
pQCD FONLL calculation~\cite{Cacciari:2005rk}.}
\label{fig_bfrac}
\end{figure}

The ratio of electrons from bottom to inclusive heavy-flavor decays, 
$b\rightarrow e / (c\rightarrow e{\rm } + {\rm } b\rightarrow e)$, can 
be constructed from the electron cross sections, and is shown in 
Fig.~\ref{fig_bfrac_fonll}. In this measurement, the electrons from the 
feed-down decay $b\rightarrow c \rightarrow e$ are considered part of 
the bottom electron sample. The contribution of bottom decays to the 
inclusive electron sample is seen to increase sharply with $p_T$, coming 
to dominate over that of charm quarks above $p_T\approx 4$ GeV/$c$. The 
solid gray line corresponds to the FONLL calculation, with its 
uncertainty depicted by dashed gray lines. The measured bottom electron 
fraction is observed to be consistent with the FONLL calculation within 
uncertainties. In particular, good agreement with the central FONLL 
value is seen below $p_T\approx 3$ GeV/$c$, with the measured fraction 
rising slightly above that at higher $p_T$.

Figure~\ref{fig_bfrac} shows a comparison of the unfolded bottom electron 
fraction with earlier measurements made by the 
STAR~\cite{Aggarwal:2010xp} and PHENIX~\cite{Adare:2009ic} 
collaborations using electron-hadron and electron-$D$ correlations. It 
is apparent that the size of the dataset, combined with the unfolding 
method used in the present analysis provides a result with smaller total 
uncertainty and significantly extended kinematic reach at low $p_T$. 
Furthermore, the unfolded result provides a more direct determination of 
the bottom electron fraction since---unlike the earlier 
measurements---it does not depend on model-dependent \textsc{pythia} 
templates of event kinematics to describe the shape of electron-hadron 
correlations.

The previous PHENIX electron-hadron results, plotted with blue markers, 
are in good agreement with the new unfolded measurement. Similarly, the 
STAR measurement---while systematically lower---is also consistent with 
the unfolded bottom fraction on account of its large combined 
statistical and systematic uncertainties. The degree of agreement 
between these two measurements was quantified by calculating---under the 
null hypothesis that the two results are equal---the probability of 
obtaining a difference more extreme than that currently observed between 
the measurements. The resulting $p$-value using a two-sample chi-square 
statistic is found to be 0.15, indicating that the null hypothesis 
cannot be rejected. In short, both the unfold and the STAR bottom 
fraction measurements are consistent given the uncertainties.

\section{Summary}

We have reported on a new measurement of the differential-invariant 
production cross section of separated-heavy-flavor electrons in 
$p$$+$$p$ collisions at $\sqrt{s}=200$ GeV, at midrapidity $|\eta|<0.35$ 
and within $1.0 < p_T^{(e)} < 9.0$ GeV/$c$.  The measurement proceeds 
via an unfolding analysis where the yield of open-heavy-flavor hadrons 
is inferred from the inclusive-heavy-flavor electron spectrum, and the 
electron DCA$_T$ distribution measured with the PHENIX silicon-vertex 
detector. The individual yields of charm and bottom electrons, as well 
as the bottom electron fraction, are found to be consistent with FONLL 
calculations. This measurement will provide a precision baseline for 
future-heavy-flavor-separation analyses. In particular, forthcoming 
PHENIX results using a high-statistics Au$+$Au dataset promise to reduce 
current uncertainties and shed light on the centrality dependence of 
charm and bottom suppression.

%%%%%%%%%%%%%%%%%%%%%%%%%%%%%%%%%%%%%%%%%%%%%%%%%%%%%%  Acknowledgements

%\section*{ACKNOWLEDGMENTS}   % MGS17 long form for all journals

%%%%%%%%%%%%%%%%%%%%%%  ACKNOWLEDGMENTS}  %%%%% MGS19 version
%% 2018 change in Korea

\begin{acknowledgments}

We thank the staff of the Collider-Accelerator and Physics
Departments at Brookhaven National Laboratory and the staff of
the other PHENIX participating institutions for their vital
contributions.  We acknowledge support from the
Office of Nuclear Physics in the
Office of Science of the Department of Energy,
the National Science Foundation,
Abilene Christian University Research Council,
Research Foundation of SUNY, and
Dean of the College of Arts and Sciences, Vanderbilt University
(U.S.A),
Ministry of Education, Culture, Sports, Science, and Technology
and the Japan Society for the Promotion of Science (Japan),
Conselho Nacional de Desenvolvimento Cient\'{\i}fico e
Tecnol{\'o}gico and Funda\c c{\~a}o de Amparo {\`a} Pesquisa do
Estado de S{\~a}o Paulo (Brazil),
Natural Science Foundation of China (People's Republic of China),
Croatian Science Foundation and
Ministry of Science and Education (Croatia),
Ministry of Education, Youth and Sports (Czech Republic),
Centre National de la Recherche Scientifique, Commissariat
{\`a} l'{\'E}nergie Atomique, and Institut National de Physique
Nucl{\'e}aire et de Physique des Particules (France),
Bundesministerium f\"ur Bildung und Forschung, Deutscher Akademischer
Austausch Dienst, and Alexander von Humboldt Stiftung (Germany),
J. Bolyai Research Scholarship, EFOP, the New National Excellence
Program ({\'U}NKP), NKFIH, and OTKA (Hungary),
Department of Atomic Energy and Department of Science and Technology
(India),
Israel Science Foundation (Israel),
Basic Science Research and SRC(CENuM) Programs through NRF
funded by the Ministry of Education and the Ministry of
Science and ICT (Korea).
Physics Department, Lahore University of Management Sciences (Pakistan),
Ministry of Education and Science, Russian Academy of Sciences,
Federal Agency of Atomic Energy (Russia),
VR and Wallenberg Foundation (Sweden),
the U.S. Civilian Research and Development Foundation for the
Independent States of the Former Soviet Union,
the Hungarian American Enterprise Scholarship Fund,
the US-Hungarian Fulbright Foundation,
and the US-Israel Binational Science Foundation.

\end{acknowledgments}

%\clearpage

%%%%%%%%%%%%%%%%%%%%%%%%%%%  References

%\bibliography{ppg223x1} 

 
\end{document}